%% file: Vision-based_130707.tex
\documentclass[12pt]{article}
\usepackage[english]{babel}
\usepackage[latin1]{inputenc}
\usepackage{amsfonts,amssymb,amsmath, epsfig}
\usepackage{color,graphicx,graphics,psfrag}
\usepackage{amsmath,amstext,amssymb,amsfonts, amscd}
\usepackage{hyperref}           

\def\Box{\leavevmode\vbox{\hrule
     \hbox{\vrule\kern4pt\vbox{\kern4pt}%
           \vrule}\hrule}}
\def\blackbox{\leavevmode\vrule height 5pt width 4pt depth 0pt\relax}
\def\endproof{\null\hfill {$\blackbox$}\bigskip}

\newcounter{appendix}
\def\appendix{\advance\c@appendix by 1
   \def\thesection{\Alph{section}}
   \ifnum\c@appendix=1 \setcounter{section}{-1} \fi
   \@startsection {section}{1}{\z@}{-3.5ex plus -1ex minus 
   -.2ex}{2.3ex plus .2ex}{\Large\bf}}

\def\paragraph#1{{\bf #1\ }}

\newtheorem{lemma}{Lemma}[section]

\newtheorem{remark}{Remark}[section]
\title{Vision-based macroscopic pedestrian models} 
\author{P. Degond $^{(1,2)}$, C. Appert-Rolland$^{3,4}$, J. Pettre$^{5}$, G. Theraulaz$^{6,7}$} 
\date{} 
\begin{document}

\maketitle

\vspace{0.5 cm}

\begin{center}
1-Université de Toulouse; UPS, INSA, UT1, UTM ;\\ 
Institut de Mathématiques de Toulouse ; \\
F-31062 Toulouse, France. \\
2-CNRS; Institut de Mathématiques de Toulouse UMR 5219 ;\\ 
F-31062 Toulouse, France.\\
email: pierre.degond@math.univ-toulouse.fr
\end{center}

\begin{center}
3- Laboratoire de Physique Théorique, Université Paris Sud, \\
bâtiment 210, 91405 Orsay cedex, France\\
4- CNRS, UMR 8627, Laboratoire de physique théorique, 91405 Orsay, France\\
email: Cecile.Appert-Rolland@th.u-psud.fr
\end{center}

\begin{center}
5-INRIA Rennes - Bretagne Atlantique, Campus de Beaulieu, 35042 Rennes, France \\
email: julien.pettre@irisa.fr
\end{center}

\begin{center}
6-Centre de Recherches sur la Cognition Animale, UMR-CNRS 5169, \\ Université Paul Sabatier, Bât 4R3, \\
118 Route de Narbonne, 31062 Toulouse cedex 9, France. \\
7- CNRS, Centre de Recherches sur la Cognition Animale, F-31062 Toulouse, France \\
email: theraula@cict.fr
\end{center}

\vspace{0.5 cm}
\begin{abstract}
We propose a hierarchy of kinetic and macroscopic models for a system consisting of a large number of interacting pedestrians. The basic interaction rules are derived from \cite{Ondrej_etal_Siggraph10} where the dangerousness level of an interaction with another pedestrian is measured in terms of the derivative of the bearing angle (angle between the walking direction and the line connecting the two subjects) and of the time-to-interaction (time before reaching the closest distance between the two subjects). A mean-field kinetic model is derived. Then, three different macroscopic continuum models are proposed. The first two ones rely on two different closure assumptions of the kinetic model, respectively based on a  monokinetic  and a von Mises-Fisher distribution. The third one is derived through a hydrodynamic limit. In each case, we discuss the relevance of the model for practical simulations of pedestrian crowds.
\end{abstract}

\medskip
\noindent
{\bf Acknowledgments:} This work has been supported by the French 'Agence Nationale pour la Recherche (ANR)' in the frame of the contracts 'Pedigree' (ANR-08-SYSC-015-01) and 'CBDif-Fr' (ANR-08-BLAN-0333-01)

\medskip
\noindent
{\bf Key words: } Pedestrian dynamics, Synthetic-vision, Individual-Based Models, Kinetic Model, Fluid model, Closure relation, monokinetic, von Mises-Fisher distribution, Local Thermodynamic Equilibrium, hydrodynamic limit

\medskip
\noindent
{\bf AMS Subject classification: } 35L60, 35L65,, 76Z99, 91A13, 91F99
\vskip 0.4cm

%%%%%%%%%%%%%%%%%%%%%%%%%%%%%%%%%%%%%%%%%%%%%%%%
%%%%%%%%%%%%%%%%%%%%%%%%%%%%%%%%%%%%%%%%%%%%%%%%
%%%%%%%%%%%%%%%%%%%%%%%%%%%%%%%%%%%%%%%%%%%%%%%%
%%%%%%%%%%%%%%%%%%%%%%%%%%%%%%%%%%%%%%%%%%%%%%%%
\setcounter{equation}{0}
\section{Introduction}
\label{sec_intro}

Crowd simulation has become a major subject of research with a wide field of applications. These include safety for preventing crowd disasters or assessing evacuation scenarios, architecture  for assessing the level of service of public spaces, or urban design for ensuring traffic efficiency and comfort. In spite of these very important issues, the status of research in crowd dynamics is still rather preliminary and is certainly less mature than the comparable field of vehicular traffic.

In this paper, we propose a hierarchy of models derived from the vision-based Individual-Based Model (IBM) of \cite{Ondrej_etal_Siggraph10}. In this model, the pedestrians make an assessment of the dangerousness of an encounter with the other pedestrians and take decisions of turning or slowing down based on this assessment. To make this assessment, each pedestrian senses the Derivative of the Bearing Angle (DBA) and the Time-To-Interaction (TTI) of each of his interaction partners. The bearing angle is the angle between the walker's direction and the line connecting the two interaction partners. Its time derivative provides a sensor of the likeliness of a collision,  a constant bearing angle in time being associated with a risk of future collision. The TTI is the time to closest approach between the pedestrians assuming linear motion. Combined DBA and TTI information gives access to an assessment of the risk of collisions in the future, and how imminent these collisions may be. The psychological literature shows that these two quantities can be visually perceived by the pedestrian from his optical flow  \cite{Cutting_etal_PsychRev95}.  The assumption that the subjects control their trajectories based on an evaluation of these indicators has solid grounds \cite{Vandenberg_Overmars_IntJRoboticsRes08}. In addition, the IBM model of \cite{Ondrej_etal_Siggraph10} suggests that agents deviate in order to avoid future collision while they decelerate to avoid imminent collision.

Specifically, in this paper, we propose a Kinetic Model (KM) and three Continuum Models (CM) derived from the IBM of \cite{Ondrej_etal_Siggraph10}. The CM are respectively based on monokinetic and von Mises-Fisher (VMF) closures for the first two ones, and on a hydrodynamic limit for the third one. We apply the same methodology as in a previous work \cite{Degond_etal_Heuristics}. In this reference, following \cite{Moussaid_etal_PNAS11}, the decision-making phase takes the form of a game where the subjects choose the local optimal trajectory towards their goal while avoiding collisions with the other pedestrians. Here, the decision-making phase is based on avoiding collision-threatening situations. From the cognitive viewpoint, it is not clear which of the mechanisms of \cite{Moussaid_etal_PNAS11} or \cite{Ondrej_etal_Siggraph10} is more realistic. The collision sensors are different: they are made of the pair (TTI, DBA) in \cite{Ondrej_etal_Siggraph10} and by the pair (TTI, DTI) in \cite{Moussaid_etal_PNAS11}, where the Distance-To-Interaction (DTI) is the distance to the point where a collision is anticipated, assuming that each pedestrian moves at a constant speed. The DBA used in \cite{Ondrej_etal_Siggraph10} is a variable which is directly perceived by the pedestrian from his visual field. The DTI is estimated from the knowledge of the velocities of the pedestrians, but the model of \cite{Moussaid_etal_PNAS11} does not invoke the way by which these velocities are estimated. In spite of their different hypotheses, these two approaches lead to fairly similar macroscopic models, as will be seen in the present paper, by comparison with \cite{Degond_etal_Heuristics}. The models of \cite{Degond_etal_Heuristics} and of the present paper have the same mathematical structure and they only differ through the modeling details of the interactions. 

The force term used by  \cite{Moussaid_etal_PNAS11} has a potential structure, which results from the game-theoretic  framework of the model, and which is extensively used in \cite{Degond_etal_Heuristics}. By contrast, as we will see in the present paper, the model of \cite{Ondrej_etal_Siggraph10} does not have such a potential structure and leads to more complex macroscopic models. However, the game-theoretic framework of \cite{Moussaid_etal_PNAS11} can be easily translated to the model of \cite{Ondrej_etal_Siggraph10} by slightly modifying the avoidance rules. We refer to this modification as the 'potential-driven dynamics'. This modification will allows us to implement the same type of methodologies as in \cite{Degond_etal_Heuristics}. 

Indeed, while the first two fluid closures, the monokinetic and VMF closures are implementable in the original framework of \cite{Ondrej_etal_Siggraph10}, the third one, based on the hydrodynamic limit, requires the use of the potential-driven dynamic in a spatially local approximation. It relies on the use of a Local Thermodynamic Equilibrium (LTE) which can be viewed as a Nash equilibrium for a game using this potential as a cost function. This is a special example of a general framework relating game theory and kinetic theory which has been proposed in \cite{Degond_etal_arXiv:1212.6130} and which bears analogies with the so-called Mean-Field Games \cite{Lasry_Lions_JapanJMath07}. This framework has been applied to the context of pedestrians for the first time in \cite{Degond_etal_Heuristics}. 

A recent review on crowd modeling can be found in \cite{Bellomo_Dogbe_SIAMRev11}. Crowd simulation models are mostly built on IBM \cite{Helbing_BehavSci91, Helbing_Molnar_PRE95, Helbing_Molnar_SelfOrganization97, Paris_etal_Eurographics07, Reynolds_Siggraph87, Reynolds_ProcGameDev99, Shao_Terzopoulos_Siggraph05, Vandenberg_etal_RoboticsResearch11} or on cellular automata \cite{Nishinari_etal_IEICETranspInfSyst04}. The model discussed here \cite{Ondrej_etal_Siggraph10} belongs to the class of vision-based models, which describe the response of the subjects to the visual scene in front of them. According to various types of stimuli, pedestrians anticipate the occurrence of collisions with partners and decide to turn away in response to the most threatening ones \cite{Guy_etal_Siggraph09, Paris_etal_Eurographics07, Pettre_etal_Siggraph09, Vandenberg_Overmars_IntJRoboticsRes08}. In \cite{Ondrej_etal_Siggraph10}, the formulation of the collision threat is made in terms of variables that are more immediately accessible to pedestrians through the analysis of their visual field, namely the TTI and the DBA. Once the collision threat is evaluated, the pedestrians perform an optimisation in order to avoid collisions while keeping close to their desired trajectory. Several types of optimisations can be performed, and it is not clear yet which one is the most relevant \cite{Guy_etal_PRE12, Hoogendoorn_Bovy_OptControlApplMeth03, Moussaid_etal_PNAS11}. The differences between the models is more thoroughly discussed in section \ref{subsec_Nped}. Other IBM's are based on traffic models \cite{Lemercier_etal_Eurographics12}. 

CM have been pioneered by \cite{Henderson_TranspRes74} and the link to the underlying IBM, explored in \cite{AlNasur_Kachroo_IEEEITSC06, Burger_etal_KRM11, Chertock_etal_preprint12, Cristiani_etal_MMS11, Helbing_ComplexSyst92, Treuille_etal_Siggraph06}. Direct derivation of CM from optimization rules can be developed \cite{Huang_etal_TranspResB09, Hughes_TranspResB02, Hughes_AnnRevFluidMech03, Jiang_etal_PhysicaA10}. The analogy with car traffic has also been extensively developed \cite{Appert-Rolland_etal_NHM11, Bellomo_Dogbe_M3AS08, Berres_etal_NHM11, Colombo_Rosini_MMAS05, Coscia_Canavesio_M3AS08, Motsch_etal_friction, Piccoli_Tosin_ContMechThermo09}. One difficulty with crowd simulations at very high densities is  the handling of the volume exclusion constraint. This specific question has been investigated in several references \cite{Degond_Hua_arXiv:1207.3522, Degond_etal_JCP11, Maury_etal_NHM11, Narain_etal_Siggraph09}. The mathematical theory of some crowd CM's has been initiated in \cite{DiFrancesco_etal_JDE11}. KM, which are intermediate between IBM and CM have not received much attention in the context of crowd modeling so far \cite{Bellomo_Dogbe_SIAMRev11, Bellomo_Bellouquid_MathModelCollectivBehav10}.

The outline of the paper is as follows. The IBM of \cite{Ondrej_etal_Siggraph10} is reviewed in section \ref{sec:IBM}. The passage to the subsequent KM is performed in section \ref{sec_mean_field}. The derivation of the continuum models is realized in section \ref{sec:macro}. The obtained models are discussed mostly in reference to \cite{Degond_etal_Heuristics} in section \ref{sec:discussion} (a thorough discussion with the literature can be found in \cite{Degond_etal_Heuristics}). Finally, a conclusion is drawn in section \ref{sec:conclu}.

%%%%%%%%%%%%%%%%%%%%%%%%%%%%%%%%%%%%%%%%%%%%%%%%
%%%%%%%%%%%%%%%%%%%%%%%%%%%%%%%%%%%%%%%%%%%%%%%%
%%%%%%%%%%%%%%%%%%%%%%%%%%%%%%%%%%%%%%%%%%%%%%%%
%%%%%%%%%%%%%%%%%%%%%%%%%%%%%%%%%%%%%%%%%%%%%%%%
\setcounter{equation}{0}
\section{The vision-based Model of pedestrian motion}
\label{sec:IBM}

%%%%%%%%%%%%%%%%%%%%%%%%%%%%%%%%%%%%%%%%%%%%%%%%
%%%%%%%%%%%%%%%%%%%%%%%%%%%%%%%%%%%%%%%%%%%%%%%%
\subsection{collision perception phase}
\label{sec:collision_perception}

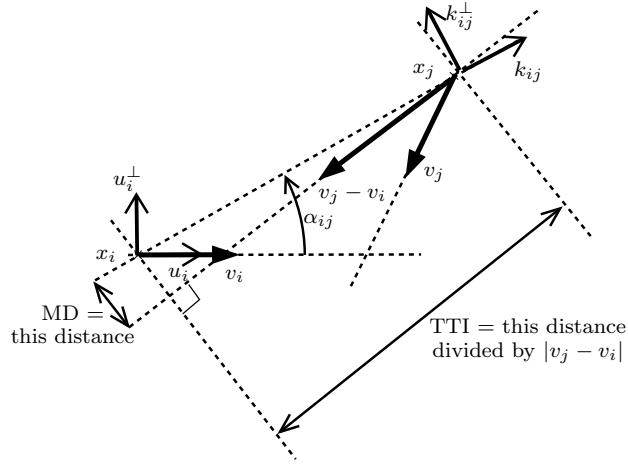
\begin{figure}
\begin{center}
\input{collision_1_1_2.pstex_t}
\caption{Geometry of a collision. The MD is the smallest distance which separates the two pedestrians $i$ and $j$ supposing that they cruise on a straight line at constant velocities $v_i$ and $v_j$. The point on pedestrian $i$'s trajectory where the minimal distance is attained is the interaction point $x_{i \, \mbox{\scriptsize{int}}}$ of pedestrian $i$ in his interaction with pedestrian $j$. The MD is the distance between $x_{i \, \mbox{\scriptsize{int}}}$ and $x_{j \,  \mbox{\scriptsize{int}}}$. The DTI is the distance which separates the current pedestrian position $x_i$ to the interaction point $x_{i \, \mbox{\scriptsize{int}}}$. The Time-To-Interaction (TTI) is the time needed by pedestrian $i$ to reach the interaction point from his current position. Clearly, TTI $=$ DTI$/|v_i|$. }
\label{Fig_collision_1_1_2}
\end{center}
\end{figure}

Figure \ref{Fig_collision_1_1_2} gives a schematic picture of the geometry of the interaction between two pedestrians. We consider a pedestrian $i$ located at a position $x_i(t)$, with a velocity $v_i$. He interacts with another pedestrian $j$ located at a position $x_j(t)$ who has velocity $v_j$. The first indicator of the dangerousness of the collision measured by pedestrian $i$ is the time Derivative of the Bearing Angle (DBA). The bearing angle $\alpha_{ij}$ under which pedestrian $i$ sees his collision partner $j$ is the angle between $v_i$ and $x_j - x_i$: 
$$ \alpha_{ij} = \widehat{(v_i, x_j - x_i)}. $$
The DBA $\dot \alpha_{ij}$ is the time derivative of $ \alpha_{ij}$, i.e. $\dot \alpha_{ij} = d \alpha_{ij} / dt$. Small values of $\dot \alpha_{ij}$ indicate that a collision between the two pedestrians is very likely, as shown below. If the pedestrians were point particles, their trajectory would interesect if and only if the DBA were exactly zero. Since pedestrians have a finite size, the intersection occurs for small but non-zero values of the DBA. As we will see later on, the DBA is proportional to the square of the reciprocal of the distance between the agents. Therefore, when the pedestrians are very close, collisions happen even if the DBA is fairly large.

The second 'indicator' of the dangerousness of a collision is the Time-To-Interaction (TTI) denoted by $\tau_{ij}$. We first define the interaction point as the point on the subject's future trajectory where the distance to his collision partner is minimal. This Minimal Distance (MD) is denoted by $D_{ij}$. The TTI is the time needed by the subject to reach this interaction point from his current position. 
The TTI and MD are computed by the agents assuming that both agents move in straight lines with constant speeds i.e. that $v_i$ and $v_j$ are constant vectors in time. It is not unlikely that pedestrians are able to anticipate more complex motions such as circular motion but this aspect is left to future work.  

The expressions of the DBA and TTI are given below. The TTI can also be found in \cite{Degond_etal_Heuristics} but we recall its computation below for the sake of completeness. We introduce the following notations: for two 2-dimensional vectors $A=(a_1,a_2)$ and $B=(b_1,b_2)$, we denote by 
$$ (A \cdot B) = a_1 b_1 + a_2 b_2 = |A| \, |B| \, \cos \widehat{(A,B)}, $$ 
the scalar product of $A$ and $B$. We denote by 
$A \times B$ the vector product, which, in the present 2-dimensional context, is defined as the following scalar:
\begin{equation}
A \times B = a_1 b_2 - a_2 b_1 = (A \cdot B^\bot) = |A| \, |B| \, \sin \widehat{(A,B)} = \det \{A,B\}. 
\label{eq:vectprod}
\end{equation}
Here, $B^\bot = (b_2, -b_1)$ is the vector obtained from $B$ by a rotation of an angle $\pi/2$, 'det' denotes the determinant, and $\{A,B\}$ is the matrix whose columns are respectively $A$ and $B$. 

\begin{lemma}
We have (see Fig. \ref{Fig_collision_1_1_2}): 
\begin{eqnarray}
& & \hspace{-1cm}
\dot \alpha_{ij} = \frac{(x_j-x_i) \times (v_j-v_i)}{|x_j-x_i|^2},  
\label{eq:bearing} \\
& & \hspace{-1cm}
\tau_{ij} = - \frac{(x_j-x_i) \cdot (v_j-v_i)}{|v_j-v_i|^2},  
\label{eq:tti} \\
& & \hspace{-1cm}
D_{ij} = \Big( |x_j - x_i|^2 - \big( (x_j - x_i) \cdot \frac{v_j - v_i}{|v_j - v_i|} \big)^2 \Big)^{1/2}. 
\label{eq:min_dist}
\end{eqnarray}
\label{lem:bearing}
\end{lemma}

\medskip
\noindent
{\bf Proof.}  We introduce the unit vector $k_{ij}$ of the line connecting the two agents, the relative velocity $v_{ij}$ of agent $j$ relative to $i$ and the distance $d_{ij}$ between the agents. These quantities are defined by the following relations:
\begin{equation} d_{ij}(t) = |x_j(t) - x_i(t)| , \quad k_{ij}(t) = \frac{x_j(t) - x_i(t)}{d_{ij}(t)}, \quad  v_{ij} = v_j - v_i .  
\label{eq:defk}
\end{equation}
We also introduce the unit vector $k_{ij}^\bot$, orthogonal to $k_{ij}$ and such that $(k_{ij},k_{ij}^\bot)$ is a direct ortho-normal frame.  We denote by $u_i = v_i/|v_i|$ the unit vector in the direction of $v_i$ and define its orthonormal complement $u_i^\bot$ which is a unit vector making the pair $(u_i,u_i^\bot)$ a direct ortho-normal frame (see Fig. \ref{Fig_collision_1_1_2}). 

Now, by the definition of the bearing angle $\alpha_{ij}$, we can write: 
$$ k_{ij} = u_i \cos \alpha_{ij} + u_i^\bot \sin \alpha_{ij}. $$
Taking the time derivative of this relation and using the fact that $u_i$ and $u_i^\bot$ are constant (since the motion of pedestrian $i$ is supposed rectilinear with constant speed) leads to 
$$ \dot k_{ij} = \dot \alpha_{ij} \, (- u_i \sin \alpha_{ij} + u_i^\bot \cos \alpha_{ij}) = \dot \alpha_{ij} k_{ij}^\bot. $$
On the other hand, taking the time derivative of the first equation (\ref{eq:defk}), and after some easy computations, we find 
\begin{eqnarray*}
\dot k_{ij} (t) &=&  \frac{d}{dt} \Big( \frac{x_j(t) - x_i(t)}{d_{ij}(t)} \Big) \\
&=& \frac{1}{d_{ij}(t)} \big( v_{ij} - ( v_{ij} \cdot k_{ij}(t) ) \,  k_{ij}(t) \big) \\
&=& \frac{1}{d_{ij}(t)} \, ( v_{ij} \cdot k_{ij}^\bot(t) ) \, k_{ij}^\bot(t) .
\end{eqnarray*}
Identifying these two relations and using (\ref{eq:vectprod}), we get
$$ \dot \alpha_{ij} = \frac{1}{d_{ij}(t)} ( v_{ij} \cdot k_{ij}^\bot(t) ) = \frac{1}{d_{ij}(t)} ( v_{ij} \times k_{ij}(t) ), $$
which gives rise to formula (\ref{eq:bearing}) for the DBA.

Now, we turn to the computation of the TTI $\tau_{ij}$ and the MD $D_{ij}$. Here, we follow the proof of \cite{Degond_etal_Heuristics}. Starting from time $t$, we compute the distance $d_{ij}(t')$ at later times $t'>t$ supposing that the motion is rectilinear with constant speed, i.e. that $v_i$ and $v_j$ are constant. The distance $d_{ij}(t')$ is given by:
\begin{eqnarray}
&& \hspace{-1cm}
d_{ij}^2(t') = |x_j + v_j t' - (x_i + v_i t')|^2 \nonumber \\
&& \hspace{-1cm}
= |v_{ij}|^2 \, \Big( t' + \frac{(x_j  - x_j) \cdot v_{ij}}{|v_{ij}|^2} \Big)^2 + |x_j  - x_j |^2  - \frac{\big( (x_j  - x_j) \cdot v_{ij} \big)^2}{|v_{ij}|^2}, 
\label{eq:D2}
\end{eqnarray} 
denoting by $x_i$ and $x_j$ the positions of the two particles at time $t$. This quadratic function of time is minimal at the time  
$t'=\tau_{ij}$ given by (\ref{eq:tti}), which gives the value of the TTI. 
Finally, the MD $D_{ij}$ is given by the minimal value of (\ref{eq:D2}), i.e. $D_{ij} = d_{ij}(\tau_{ij})$, which leads to (\ref{eq:min_dist}) and ends the proof of the Lemma. \endproof

The interaction is threatening only if the TTI is positive. Indeed, if the TTI is negative, the distance to the encounter is an increasing function of time (the squared distance being a quadratic function) and there is no threat of collision in the future times. The TTI is positive if and only if $(x_j-x_i) \cdot (v_j-v_i) <0$. Furthermore, if the MD is larger than a certain threshold $R$ identified as the diameter of the individuals, plus a certain safe-keeping distance, the interaction is no longer perceived as a collision threat. Therefore, there is no interaction unless both following conditions are simultaneously satisfied: 
\begin{eqnarray}
& & \hspace{-1cm}
\tau_{ij} \geq 0 \quad \mbox{ and } \quad  D_{ij} \leq R . 
\label{eq:interaction_cnd}
\end{eqnarray}

\begin{remark}
In view of (\ref{eq:bearing}) and (\ref{eq:min_dist}) the DBA can be related to the MD by
$$ |\dot \alpha_{ij}| = \frac{|v_{ij}|}{d_{ij}^2} D_{ij} . $$
Therefore, if there is a collision threat, i.e. if $ D_{ij} \leq R$, then $|\dot \alpha_{ij}| \leq \frac{|v_{ij}|}{d_{ij}^2} \, R$. Consequently, collision threatening situations are associated to small DBA's. The collision avoidance manoeuver consists in turning to increase the magnitude of the DBA as we will see in the next section. 
\label{rem:relation_TTI_DBA}
\end{remark}

%%%%%%%%%%%%%%%%%%%%%%%%%%%%%%%%%%%%%%%%%%%%%%%%
%%%%%%%%%%%%%%%%%%%%%%%%%%%%%%%%%%%%%%%%%%%%%%%%
\subsection{Decision-making phase}
\label{sec:decision_making_OPOD}

The decision-making model of \cite{Ondrej_etal_Siggraph10} is made of two components: collision avoidance on the one hand and satisfaction of the goal on the other hand. We will add a third component, namely, noise, in order to take into account some uncertainty and variability of the subjects' responses to a given situation. We successively examine these various aspects.

%%%%%%%%%%%%%%%%%%%%%%%%%%%%%%%%%%%%%%%%%%%%%%%%
\subsubsection{Collision avoidance}
\label{subsec:collision_avoidance}

In the collision avoidance model of \cite{Ondrej_etal_Siggraph10}, the agents have two control variables, their direction of motion and their speed.  The agents avoid ``future collisions", defined by a moderate positive value of the TTI and a low value of the DBA,  by turning, i.e. changing their direction of motion. However, the ``imminent collisions", defined by a low positive value of the TTI are avoided by slowing down. Here, we make the assumption that all pedestrians move with constant speed equal to $c$. This assumption is made for simplicity only and will be waived in future work. Thus, we discard the speed as a control variable, and consequently, we assume that the imminent collisions are scarce. Indeed, acting on the direction of motion only is not sufficient to prevent imminent collisions (as the centripetal force that the pedestrians are able to develop in order to turn has an upper bound related to their muscular capacity). Therefore, the present constant speed model cannot completely rule out the fact that pedestrians might actually interpenetrate each other, which is obviously unrealistic. This implicitly restricts the model to low density crowds, where the imminent collisions are less likely. By ignoring the speed as a control variable, we simplify the model and allow us to focus on the directional changes only. Consequently, in the remainder of the paper, we suppose that $|v_i|=|v_j|=c$. Then, we have $v_i = c \, u_i$, $v_j = c \, u_j$ with $|u_i| = |u_j| = 1$. We will also assume that there are no fixed obstacles and thus, the only obstacles consist of other pedestrians. 

In this section, we describe the response of pedestrian $i$ to the perception of the DBA $\dot \alpha_{ij}$ and the TTI $\tau_{ij}$ of a single other pedestrian $j$. The model follows the lines of \cite{Ondrej_etal_Siggraph10}, with some simplifications of the expressions of the collision avoidance response. We assume that pedestrian $i$ reacts to the likeliness of a collision with $j$ by rotating with angular velocity $\omega_{ij}$, and similarly for his collision partner $j$, with an angular velocity $\omega_{ji}$. We note that the collision indicators, $\dot \alpha_{ij}$ and $\tau_{ij}$ are the same for $i$ and $j$. Therefore, we expect that the responses of the two collision partners to be symmetric, i.e. $\omega_{ji} = \omega_{ij}$. This leads to the following equations of motion: 
\begin{eqnarray}
& & \hspace{-1cm} 
\dot x_i = c \, u_i, \qquad  \qquad  \dot u_i = \omega_{ij} u_i^\bot, 
\label{eq:dotx} \\ 
& & \hspace{-1cm} 
\dot x_j = c \, u_j, \qquad  \qquad \dot u_j = \omega_{ij} u_j^\bot. 
\label{eq:dotu}
\end{eqnarray}

We now establish the expression $\omega_{ij}$, following \cite{Ondrej_etal_Siggraph10}. First, the fact that there is no interaction unless condition (\ref{eq:interaction_cnd}) is satisfied implies that $\omega_{ij}$ involves a factor $H (\tau_{ij}) \, H(R^2 - D_{ij}^2)$, where $H$ is the Heaviside function (i.e. the indicator function of the set of positive real numbers). The second observation is that collision avoidance is obtained by increasing the magnitude of the DBA $\dot \alpha_{ij}$. So, $\omega_{ij}$ must have an opposite sign to $\dot \alpha_{ij}$. Finally, $\omega_{ij}$ must increase when the risk of collision increases, through a function $\Phi(|\dot \alpha_{ij}|, |\tau_{ij}|) \geq 0$ to be determined below. Therefore, we can write
\begin{eqnarray}
& & \hspace{-1cm} 
\omega_{ij} = - \mbox{Sign}(\dot \alpha_{ij}) \, H(\tau_{ij}) \, H(R^2 - D_{ij}^2) \,  \Phi(|\dot \alpha_{ij}|, |\tau_{ij}|) \,
.
\label{eq:omega}
\end{eqnarray}
In the next lemma, we verify that this expression tends to increase the DBA and therefore, decreases the likeliness of the collision, as seen in remark \ref{rem:relation_TTI_DBA}. 

\begin{lemma}
Suppose that $R^2 - D_{ij}^2 \geq 0$ and $\tau_{ij} >0$ and that pedestrians $i$ and $j$ follow the dynamic (\ref{eq:dotx}), (\ref{eq:dotu}) with $\omega_{ij}$ given by (\ref{eq:omega}). Then, there exists a function $\lambda_{ij}(t) \geq 0$ such that the DBA $\dot \alpha_{ij}$ between the two pedestrians satisfies: 
\begin{eqnarray}
\ddot \alpha_{ij} = c \lambda_{ij} \, \dot \alpha_{ij}.  
\label{eq:dot_alpha} 
\end{eqnarray}
As a consequence of the nonnegativity of $\lambda_{ij}(t)$, the function $|\dot \alpha_{ij}(t)|$ is increasing with time. 
\label{lem:dba_increases}
\end{lemma}

\medskip
\noindent
{\bf Proof.} We can write, using eqs. (\ref{eq:bearing}), (\ref{eq:dotx}), (\ref{eq:dotu}), (\ref{eq:omega}): 
\begin{eqnarray*}
c^{-1} \ddot \alpha_{ij} &=& \frac{(x_j - x_i) \times (\dot u_j - \dot u_i)}{|x_j - x_i|^2}  \\
& & \hspace{3cm} 
- 2c ((x_j - x_i) \cdot (u_j - u_i)) \frac{(x_j - x_i) \times (u_j - u_i)}{|x_j - x_i|^4}  \\
&=&  
 \frac{\omega_{ij}}{|x_j - x_i|^2}  ((x_j - x_i) \times ( u_j^\bot - u_i^\bot)) + 2 c \, \frac{|u_j - u_i|^2}{|x_j - x_i|^2} \,  \tau_{ij} \,  \dot \alpha_{ij} \\
&=& \frac{\omega_{ij}}{|x_j - x_i|^2}  ((x_j - x_i) \cdot (u_j - u_i)) + 2 c \, \frac{|u_j - u_i|^2}{|x_j - x_i|^2} \,  \tau_{ij} \,  \dot \alpha_{ij} \\
&=& c \frac{|u_j - u_i|^2}{|x_j - x_i|^2} \, \tau_{ij}  ( - \omega_{ij} + 2  \dot \alpha_{ij} ) \\
&=& c \frac{|u_j - u_i|^2}{|x_j - x_i|^2} \tau_{ij}  \left( \frac{1}{|\dot \alpha_{ij} |}  \Phi(|\dot \alpha_{ij}|, |\tau_{ij}|)  + 2 \right) \,    \dot \alpha_{ij} \\
&=& \lambda_{ij} \, \dot \alpha_{ij},   
\end{eqnarray*}
with 
\begin{eqnarray*}
\lambda_{ij} = c \frac{|u_j - u_i|^2}{|x_j - x_i|^2} \tau_{ij}  \left( \frac{1}{|\dot \alpha_{ij} |}  \Phi(|\dot \alpha_{ij}|, |\tau_{ij}|)  + 2 \right) >0 ,    
\end{eqnarray*}
which shows (\ref{eq:dot_alpha}). The second statement is obvious. \endproof

Now, we specify the function $\Phi$. In \cite{Ondrej_etal_Siggraph10}, the following form is proposed: 
\begin{eqnarray}
& & \hspace{-1cm} 
\Phi(|\dot \alpha_{ij}|, |\tau_{ij}|) = \Phi_0 \, \max \{  \sigma(|\tau_{ij}|) - |\dot \alpha_{ij}|, \, 0 \}, 
\label{eq:Phi}
\end{eqnarray}
with 
\begin{eqnarray}
& & \hspace{-1cm} 
\sigma(|\tau_{ij}|) = a + \frac{b}{(|\tau_{ij}|+\tau_0)^c}, 
\label{eq:sigma}
\end{eqnarray}
where $\phi_0$, $\tau_0$, $a$, $b$, $c$ are positive constants. In \cite{Ondrej_etal_Siggraph10}, $\phi_0=1$ and $\tau_0 = 0$ are used. The constants $a=0$, $b=0.6$, and $c=1.5$ have been determined in \cite{Ondrej_etal_Siggraph10} from fitting against experimental data. Formula (\ref{eq:Phi}) states that if the DBA is larger than a certain threshold $\sigma$, there is no threat of collision and $\Phi$ is set to zero. On the other hand, if the DBA is smaller than this threshold, the subject turns at an angular speed which is proportional to the difference between this threshold and the actual DBA. The constant $\Phi_0$ is the proportionality constant. Now, the threshold depends on the TTI and gets larger as the TTI becomes smaller. The reason is that the range of DBA's which are felt as a threat increases as the TTI decreases and the available time range to perform a maneuver becomes smaller. 

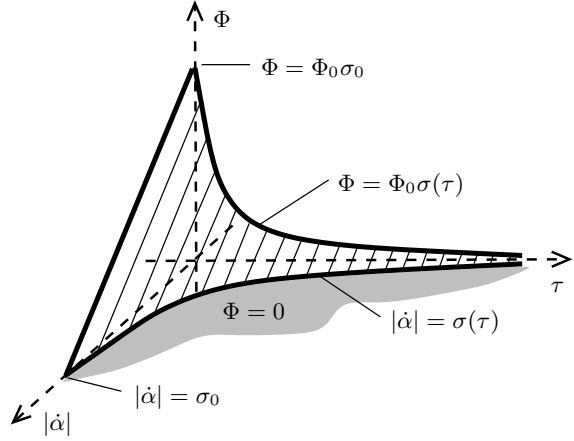
\begin{figure}
\begin{center}
\input{Phi.pstex_t}
\caption{Perspective view of the function $(|\dot \alpha|, \tau) \in [0,+\infty]^2 \to \Phi(|\dot \alpha|, \tau)$ in the case $a=0$. The value $\sigma_0$ corresponds to $\sigma_0 = \sigma(0) = \frac{b}{\tau_0^c}$ (in the case $a=0$). The function $\Phi$ is identically zero, except for the dashed area, where it is linear. The introduction of $\tau_0 >0$ makes it bounded from above by $\sigma_0$.} 
\label{fig:Phi}
\end{center}
\end{figure}

The constant $\tau_0>0$ is there to ensure that $\sigma$ and consequently $\Phi$ remain bounded, and to avoid the divergence of certain integrals in the continuum models (see section \ref{sec_mean_field}). Indeed, in realistic situations, if $\tau_{ij} \leq \tau_0$, the collision threat is such that the pedestrians not only act on the direction of their motion, but they also slow down or even stop. In this constant velocity model, we cannot take into account this feature. Therefore, we just bound the magnitude of the pedestrian angular velocity. This seems reasonable because the magnitude of the force that the pedestrians are able to exert in order to change direction is bounded by the muscular capacity. Fig. \ref{fig:Phi} provides a perspective view of the function $(|\dot \alpha|, \tau) \in [0,+\infty]^2 \to \Phi(|\dot \alpha|, \tau)$. 

To summarize, the collision avoidance model consists of eqs. (\ref{eq:dotx}), (\ref{eq:dotu}) for the pedestrian positions and velocity, together with (\ref{eq:omega}), (\ref{eq:Phi}), (\ref{eq:sigma}) for the expression of the angular velocity.

%%%%%%%%%%%%%%%%%%%%%%%%%%%%%%%%%%%%%%%%%%%%%%%%
\subsubsection{Target}
\label{subsec_target}

In their motion, the pedestrians have a goal. In \cite{Ondrej_etal_Siggraph10}, the goal is considered to be a given point, attached to each pedestrian. Navigation to the goal means that the DBA $\dot \alpha_{ig}$ of pedestrian $i$ relative to the goal $g$ should be small. The satisfaction of this constrained is realized by modifying the angular velocity (\ref{eq:omega}) in the following way:
\begin{eqnarray}
& & \hspace{-1cm} 
\omega_{ij} = \left\{ \begin{array}{l} - \mbox{Sign}(\dot \alpha_{ij}) \, H(\tau_{ij}) \, H(R^2 - D_{ij}^2) \,  \Phi(|\dot \alpha_{ij}|, |\tau_{ij}|) \, \,  \mbox{ if } \, \, |\dot \alpha_{ig}| < \Phi(|\dot \alpha_{ij}|, |\tau_{ij}|) , \\
\dot \alpha_{ig} \quad \mbox{ if } \quad  |\dot \alpha_{ig}| > \Phi(|\dot \alpha_{ij}|, |\tau_{ij}|)
. \end{array} \right.
\label{eq:omega2}
\end{eqnarray}
The rationale of this formula is as follows: We first note that (\ref{eq:omega}) gives the minimal angular velocity necessary to avoid collisions. However larger angular velocities may be chosen (bottom of (2.14)) if this allows the subject to come closer to his goal. Indeed, pedestrians reach their goal by ensuring that their DBA to the goal $\dot \alpha_{ig}$ is close to zero (in the same way that they try to avoid other pedestrians by ensuring that their DBI to their collision partner is large, as seen in section \ref{subsec:collision_avoidance}). Therefore, taking $\omega_{ij}=\dot \alpha_{ig} $ ensures that $\omega_{ij}$ has the same sign as $\dot \alpha_{ig} $ and thus that $|\dot \alpha_{ig} |$ will further decrease, as can be shown bu the same reasoning that of the proof of Lemma \ref{lem:dba_increases}. However, collision avoidance is always the priority. So, in the event where the DBI to the target is smaller than the minimal DBI needed to avoid collisions, expressed by the function $\Phi(|\dot \alpha_{ij}|, |\tau_{ij}|)$, then the latter is chosen as angular velocity $\omega_{ij}$, which is what the top relation (\ref{eq:omega2}) expreses.

%%%%%%%%%%%%%%%%%%%%%%%%%%%%%%%%%%%%%%%%%%%%%%%%
\subsubsection{Noise}
\label{subsec_noise}

In \cite{Ondrej_etal_Siggraph10}, no stochasticity is considered. However, obviously, the motion of pedestrians is not fully deterministic. When some decisions need to be made in front of several alternatives, the response of the subjects is subject-dependent. We can model this inherent uncertainty by adding a Brownian motion in velocity:
\begin{eqnarray*}
& & \hspace{-1cm} 
d u_i =  (\sqrt{2d} \circ dB^i_t) \cdot u_i^\bot, 
\end{eqnarray*}
where $\sqrt{2d}$ is the noise intensity and where $dB^i_t$ are standard white noises on the plane, which are independent from one pedestrian to another one. The circle means that the stochastic differential equation must be understood in the Stratonovich sense. This formula projects the white noise on the tangent plane to the circle $|u_i|=1$ and the integration of this stochastic differential equation generates a Brownian motion on this circle \cite{Hsu_AMS02}. This stochastic term adds up to the previous ones. Later on,  this noise term will be crucial in two of the three macroscopic closures that will be proposed: the VMF closure (section \ref{subsec:gaussian}) and the hydrodynamic limit (section \ref{subsec_hydro}). By contrast, the first closure, namely the monokinetic closure (section \ref{subsec_monokinetic}) will require zero noise.

%%%%%%%%%%%%%%%%%%%%%%%%%%%%%%%%%%%%%%%%%%%%%%%%
%%%%%%%%%%%%%%%%%%%%%%%%%%%%%%%%%%%%%%%%%%%%%%%%

\subsection{N-pedestrian model}
\label{subsec_Nped}

In this section, we now consider $N$ interacting pedestrians. The key question is how the binary encounters are combined and provide the final response of the pedestrian. In the social force model \cite{Helbing_BehavSci91, Helbing_Molnar_PRE95, Helbing_Molnar_SelfOrganization97}, the binary responses are summed up linearly. By contrast, in \cite{Ondrej_etal_Siggraph10} and the present work, the binary responses add up nonlinearly. In the present section, we review how this combination is performed in \cite{Ondrej_etal_Siggraph10}.
We now consider $N$ pedestrians with positions $(x_i(t))_{i=1,\ldots,N}$, velocity directions $(u_i(t))_{i=1,\ldots,N}$ (with $|u_i|=1$,  $\forall i=1,\ldots,N$), angular velocities $(\omega_i(t))_{i=1,\ldots,N}$ and target points $(\xi_i)_{i=1,\ldots,N}$. The target points are supposed independent of time for the sake of simplicity.

Each pedestrian is subject to the following equations of motion: 
\begin{eqnarray}
& & \hspace{-1cm} 
\dot x_i = c \, u_i, \quad \quad du_i = \omega_i \, u_i^\bot \, dt + (\sqrt{2d} \circ dB^i_t) \cdot u_i^\bot, 
\label{eq:motion_Nparticles}
\end{eqnarray}
where the white noises $dB^i_t$ are independent. To compute $\omega_i$, we define:
\begin{eqnarray}
& & \hspace{-1cm} 
\Phi_{+i} = \max_{j \, | \, \dot \alpha_{ij} > 0} \, \{ H(\tau_{ij}) \, H(R^2 - D_{ij}^2) \,  \Phi(|\dot \alpha_{ij}|, |\tau_{ij}|) \}, 
\label{eq:Phi+i} \\
& & \hspace{-1cm} 
\Phi_{-i} = \max_{j \, | \, \dot \alpha_{ij} < 0} \, \{ H(\tau_{ij}) \, H(R^2 - D_{ij}^2) \,  \Phi(|\dot \alpha_{ij}|, |\tau_{ij}|) \}, 
\label{eq:Phi-i} 
\end{eqnarray}
It should be noted that, in performing the evaluations (\ref{eq:Phi+i}), (\ref{eq:Phi-i}), all pedestrians are taken into account, i.e. there is no exclusion of a blind  zone behind the pedestrian. This feature could be easily added in the model but will be discarded here for simplicity.

We consider two cases: 
\begin{enumerate}
\item The current deviation to the goal is small. This means that $|\dot \alpha_{ig}|$ is smaller than the reaction induced by collision avoidance, i.e. 
\begin{eqnarray}
& & \hspace{-1cm} 
- \Phi_{-i} \, \leq  \, \dot \alpha_{ig}  \, \leq  \, \Phi_{+i}  . 
\label{eq:alphaig_small}
\end{eqnarray}
Then: 
\begin{eqnarray}
& & \hspace{-1cm} 
\omega_i = - \Phi_{+i} \, H \left( \big| \Phi_{-i} - |\dot \alpha_{ig}| \big| - \big| \Phi_{+i}-|\dot \alpha_{ig}| \big| \right) \nonumber \\
& & \hspace{4cm} 
+ \Phi_{-i} \, H \left( \big| \Phi_{+i}-|\dot \alpha_{ig}| \big| - \big| \Phi_{-i} - |\dot \alpha_{ig}| \big| \right),
\label{eq:omegai_case1}
\end{eqnarray}
where $H$ is again the Heaviside function. This formula states that the pedestrian determines the worst case in each direction (formulas (\ref{eq:Phi+i}) and (\ref{eq:Phi-i})) and then chooses the turning direction as the one which produces the smallest deviation to the goal (this selection is provided by the Heaviside functions in (\ref{eq:omegai_case1})). 
\item The deviation to the goal is large, i.e. 
\begin{eqnarray}
& & \hspace{-1cm} 
\dot \alpha_{ig} \, <  \,- \Phi_{-i}   \quad \mbox{ or } \quad \Phi_{+i} <  \dot \alpha_{ig}. 
\label{eq:alphaig_large}
\end{eqnarray}
Then, 
\begin{eqnarray}
& & \hspace{-1cm} 
\omega_i = \dot \alpha_{ig}.
\label{eq:omegai_case2}
\end{eqnarray}
In this case, the deviation to the goal is larger than the reaction to collisions and the decision is to restore a direction of motion more compatible to the goal. 
\end{enumerate}

We shortly discuss the analogies and differences with \cite{Moussaid_etal_PNAS11}. In both the present work and in \cite{Moussaid_etal_PNAS11}, the combination of the interactions is nonlinear. Indeed, in \cite{Moussaid_etal_PNAS11}, the minimal DTI among all the particles is computed. Here and in \cite{Ondrej_etal_Siggraph10}, the response to the most collision-threatening situation is chosen. Therefore, both involve some kind of minimization (or maximization) process. However, while the collision perception is similar in the two models, the decision-making process is different. In \cite{Moussaid_etal_PNAS11}, the pedestrians compute the best compromise between their goal and the necessity of avoiding the other pedestrians. In the present model and in \cite{Ondrej_etal_Siggraph10}, their decision is made in reaction to the dangerousness of an encounter. In this sense, it is likely that the process of \cite{Ondrej_etal_Siggraph10} is less efficient than that of \cite{Moussaid_etal_PNAS11}. Indeed, in \cite{Ondrej_etal_Siggraph10}, the successful direction might be found as the result of a succession of reactions (with possibilities of over-reactions). In \cite{Moussaid_etal_PNAS11} the resolution of the collision problem results from a geometrical reasoning based on the current situation with an extrapolation to future trajectories. However, it is not clear whether pedestrians are fully able to anticipate what is the 'best' choice, in particular, in a very crowded environment where the scene is very complex. It is more likely that they use a combination of 'intuitive' immediate reactions like reflexes in particular in the presence of unanticipated collisions, and of an 'intelligent' decision-making process based on a logical (although unconscious) analysis of what is their best route. 

In an attempt to reconcile the two viewpoints, we propose a modification of the decision-making of \cite{Ondrej_etal_Siggraph10} which, while using the same collision sensors, introduces some concept of 'optimal choice'. We develop it in the next section

%%%%%%%%%%%%%%%%%%%%%%%%%%%%%%%%%%%%%%%%%%%%%%%%
%%%%%%%%%%%%%%%%%%%%%%%%%%%%%%%%%%%%%%%%%%%%%%%%
\subsection{Modified N-pedestrian model with potential-driven dynamics}
\label{subsec_Nped_modif}

The decision making procedure described in the last section is very rough, particularly through (\ref{eq:omegai_case1}). Indeed, suppose that the level of the reactions to pedestrians coming from the left and the right are approximately the same. Then, the choice of turning towards the left or the right is very sensitive to perturbations, which results in a fairly unstable dynamics. Additionally, while the choice of one of the two possible turning directions decreases the threat of collisions with pedestrians coming from one side, it increases the threat of collisions with those coming from the other side. It is unlikely that, in real life, pedestrians make such radical choices. More likely, they try a third possibility. 

For this reason, while keeping the rationale of the model of \cite{Ondrej_etal_Siggraph10}, we modify the decision-making procedure, by introducing some optimal control idea, in the spirit of \cite{Guy_etal_PRE12, Hoogendoorn_Bovy_OptControlApplMeth03, Moussaid_etal_PNAS11}. For this purpose, we change the interpretation of the functions $\Phi_{\pm i}$ and rather view them as 'quantifiers' of what would be the optimal action. In the framework of optimal control theory, one rather speaks of cost functions. So, we will work with the negative of this quantifier. 

With this aim, we first need a generic definition of the DBA, TTI and MD. Let a particle be located at position $x \in {\mathbb R}^2$ with velocity direction $u \in {\mathbb S}^1$. Then, in its encounter with another particle located at position $y \in {\mathbb R}^2$ with velocity direction $v \in {\mathbb S}^1$, we can define the DBA $\dot \alpha (x,u,y,v)$, the TTI $\tau(x,u,y,v)$ and the MD $D(x,u,y,v)$ by: 
\begin{eqnarray}
& & \hspace{-1cm} 
\dot \alpha (x,u,y,v) = c \frac{(y-x) \times (v-u)}{|y-x|^2}, 
\label{eq:mf_alpha} \\
& & \hspace{-1cm} 
\tau (x,u,y,v) = - \frac{1}{c} \frac{(y-x) \cdot (v-u)}{|v-u|^2}, 
\label{eq:mf_tau} \\
& & \hspace{-1cm} 
D^2 (x,u,y,v) = |y-x|^2 - \Big( (y-x) \cdot \frac{v-u}{|v-u|} \Big)^2. 
\label{eq:mf_Dmin} 
\end{eqnarray}
Similarly, we define the DBA with respect to the goal $\xi$ by an analogous formula to (\ref{eq:mf_alpha}):
\begin{eqnarray}
& & \hspace{-1cm} 
\dot \alpha_g (x,u,\xi) = - c \frac{(\xi-x) \times u}{|\xi-x|^2}, 
\label{eq:mf_alphag} 
\end{eqnarray}
where we have assumed that the target point $\xi$ is immobile. Of course, we have
\begin{eqnarray*}
& & \hspace{-1cm} 
\dot \alpha_{ij}(t) = \dot \alpha (x_i(t),u_i(t),x_j(t),u_j(t)), \quad \tau_{ij}(t) = \tau (x_i(t),u_i(t),x_j(t),u_j(t)), \\
& & \hspace{-1cm} 
D_{ij}(t) = D (x_i(t),u_i(t),x_j(t),u_j(t)), \quad \dot \alpha_{ig} = \dot \alpha_g (x_i(t),u_i(t),\xi_i).
\end{eqnarray*}

We introduce the cost of undergoing collisions with other pedestrians, according to:
\begin{eqnarray}
& & \hspace{-1cm} 
\Phi_c(x,u,t) = - \max_{j} \, \Big\{ H \big(\tau (x,u,x_j(t),u_j(t)) \big) \, H \big(R^2 - D (x,u,x_j(t),u_j(t)) \big) \nonumber \\  & & \hspace{4cm} 
\Phi( \big|\dot \alpha(x,u,x_j(t),u_j(t)) \big|, \big|\tau(x,u,x_j(t),u_j(t)) \big|) \Big\}. 
\label{eq:Phi_c} 
\end{eqnarray}
We also introduce the cost of walking away from the goal $\xi$: 
\begin{eqnarray}
& & \hspace{-1cm} 
\Phi_t(x,u,\xi) = \chi \big(\dot \alpha_g (x,u,\xi) \big), 
\label{eq:Phi_t} 
\end{eqnarray}
where the function $\chi$ is large when $|\dot \alpha_g|$ is large. For instance, inspired by the function $\sigma$ (\ref{eq:sigma}), we can take:
\begin{eqnarray}
& & \hspace{-1cm} 
\chi (\dot \alpha_g) = \frac{\tilde b}{(|\dot \alpha_g| + \dot \alpha_0)^{\tilde c}}, 
\label{eq:chi} 
\end{eqnarray}
where the constants $\tilde b>0$, $\tilde c>0$ and $\dot \alpha_0$ must be calibrated by comparison with experimental data. Finally, we introduce the total cost
\begin{eqnarray}
& & \hspace{-1cm} 
\Phi(x,u,\xi,t) = \Phi_c(x,u,t) + \Phi_t(x,u,\xi). 
\label{eq:Phi_tot} 
\end{eqnarray}
The game for each pedestrian consists in minimizing his own cost, i.e. finding the optimal velocity $u_i(t)$ such that 
\begin{eqnarray}
& & \hspace{-1cm} 
u_i(t) = \mbox{arg} \min_{u \in {\mathbb S}^1} \Phi(x_i(t),u,\xi_i,t)  . 
\label{eq:Phi_min} 
\end{eqnarray}
where $\mbox{arg} \min$ denotes the velocity $u$ at which $\Phi(x_i(t),u,\xi_i,t)$ is minimum. To approach this minimum by means of a continuous process, one possibility is to use the steepest descent method, i.e. changing velocities in the direction opposite to the gradient of the cost function. Therefore, in our modified dynamic, we propose to choose $\omega_i$ proportional and opposite to the gradient of the cost function, i.e. according to the formula: 
\begin{eqnarray}
& & \hspace{-1cm} 
\omega_i u_i^\bot(t) = - k \nabla_u \Phi (x_i(t),u_i(t),\xi_i,t)  . 
\label{eq:Phi_grad} 
\end{eqnarray}
where $k$ is a constant characterizing the reaction time of the pedestrians. Note that by appropriately choosing the constants in the expression of $\Phi$, we can assume that $k=1$ without loss of generality. 

Finally, the modified IBM consists of the equations of motion (\ref{eq:motion_Nparticles}), supplemented with the expression (\ref{eq:Phi_grad}) of the force. We will refer to this IBM as the 'potential-driven dynamics'. 

This model bears analogies with the time continuous version of the model of \cite{Moussaid_etal_PNAS11} proposed in \cite{Degond_etal_Heuristics}. There are still differences in the way the pedestrians find their optimum. In this respect, the present model bears stronger analogies with \cite{Guy_etal_PRE12, Hoogendoorn_Bovy_OptControlApplMeth03} in the construction of a cost function. In \cite{Moussaid_etal_PNAS11}, the other pedestrians act as constraints, and the pedestrians find the best satisfaction of their goal subject to these constraints. Here, collision avoidance and satisfaction of the goal are treated on an equal footing by constructing the multi-target cost function (\ref{eq:Phi_tot}). Another difference from \cite{Moussaid_etal_PNAS11} is that we do not consider a blind zone, i.e. the observation region around each pedestrian is isotropic. A blind zone could be included easily,  but, similar to \cite{Moussaid_etal_PNAS11}, the motion would no longer be expressible in terms of the gradient of the potential function. Indeed, here as well as in \cite{Moussaid_etal_PNAS11} when the observation is supposed isotropic, the dynamics can be derived from a potential \cite{Degond_etal_Heuristics}. This has important consequences for the possibility of performing a hydrodynamic limit (see section \ref{subsec_hydro}).

%%%%%%%%%%%%%%%%%%%%%%%%%%%%%%%%%%%%%%%%%%%%%%%%
%%%%%%%%%%%%%%%%%%%%%%%%%%%%%%%%%%%%%%%%%%%%%%%%
%%%%%%%%%%%%%%%%%%%%%%%%%%%%%%%%%%%%%%%%%%%%%%%%
%%%%%%%%%%%%%%%%%%%%%%%%%%%%%%%%%%%%%%%%%%%%%%%%
\setcounter{equation}{0}
\section{Mean-field kinetic models}
\label{sec_mean_field}

%%%%%%%%%%%%%%%%%%%%%%%%%%%%%%%%%%%%%%%%%%%%%%%%
%%%%%%%%%%%%%%%%%%%%%%%%%%%%%%%%%%%%%%%%%%%%%%%%
\subsection{Derivation of the model}
\label{sub:meanfield_deriv}

We now formally derive a mean-field kinetic model for the particle system presented in the previous section. We introduce the probability distribution function $f(x,u,\xi,t)$ of particles of position $x \in {\mathbb R}^2$, velocity direction $u \in {\mathbb S}^1$, target point $\xi \in {\mathbb R}^2$ at time $t$. We recall that ${\mathbb S}^1$ denotes the set of vectors of ${\mathbb R}^2$ of unit norm. 
The quantity $f(x,u,\xi,t) \, dx \, du \, d\xi$ is the probability of finding pedestrians in a small physical volume $dx$ about point $x$, within an angular neighborhood $du$ of velocity direction $u$, and within a neighborhood $d\xi$ of target point $\xi$ at time $t$. The distribution function $f$ satisfies the following mean-field kinetic equation 
\begin{eqnarray}
& & \hspace{-1cm} 
\partial_t f + c u \cdot \nabla_x f + \nabla_u \cdot ( F_f \, f) = d \Delta_u f . 
\label{eq:kinetic}
\end{eqnarray}
The operator at the left hand-side of (\ref{eq:kinetic}) describes the motion of particles at velocity $c \, u$ and their acceleration by the force $F_f$ (which depends on $f$ itself). The diffusion operator at the right-hand side comes from the  noise. Let $\theta$  be the angle between $u$ and the first coordinate direction. Then, $ u = (\cos \theta,\sin \theta)$, $u^\bot = (-\sin \theta,\cos \theta)$, and 
$$ u \cdot \nabla_x f = \cos \theta \, \partial_{x_1} f + \sin \theta \, \partial_{x_2} f, \quad \nabla_u \cdot ( F_f \, f) = \partial_{\theta} (\omega_f f), \quad \Delta_u f = \partial^2_\theta f, $$
where 
\begin{eqnarray}
& & \hspace{-1cm} 
F_f(x,u,\xi,t) = \omega_f(x,u,\xi,t) \,  u^\bot , 
\label{eq:mf_force}
\end{eqnarray}
and $\omega_f$ is a scalar quantity, to be determined below. Because the velocity $u$ is of constant norm ($|u|=1$), the force term $F_f$ is orthogonal to $u$, i.e. is a vector proportional to $u^\bot$, as expressed by (\ref{eq:mf_force}). There is no operator acting on the $\xi$-dependence of $f$. This is because the target point $\xi$ is a fixed quantity attached to each pedestrian which does not change in time. In the case of a given external force field $F$, eq. (\ref{eq:kinetic}) just follows from the stochastic particle system (\ref{eq:motion_Nparticles}) by the application of Ito's formula. In the case of a self-consistent force field such as the one given by (\ref{eq:omegai_case1}), (\ref{eq:omegai_case2}), the rigorous derivation is an open problem (see e.g. \cite{Bolley_etal_AML12} for the derivation of a mean-field model in a different but related context). 

To find an expression for $\omega_f$, we first recall the expressions (\ref{eq:mf_alpha}), (\ref{eq:mf_tau}), (\ref{eq:mf_Dmin}) and (\ref{eq:mf_alphag}) of the DBA, TTI, MD and DBA relative to the target, respectively denoted by $\dot \alpha (x,u,y,v)$, $\tau(x,u,y,v)$, $D(x,u,y,v)$ and $\dot \alpha_g (x,u,\xi)$. Now, we have to define the analogs of $\Phi_{+i}$ and $\Phi_{-i}$ for a continuum of particles. Litterally, eqs. (\ref{eq:Phi+i}), (\ref{eq:Phi-i}) should be transformed into
\begin{eqnarray*}
& & \hspace{-1cm} 
\Phi_\pm(x,u,t) = \max_{(y,v) \in \mbox{{\scriptsize Supp}} (f(t)) \, | \, \pm \dot \alpha(x,u,y,v) > 0} \, \{ H(\tau (x,u,y,v)) \, H(R^2 - D(x,u,y,v)) \\
& & \hspace{8cm} 
  \Phi(|\dot \alpha (x,u,y,v)|, |\tau(x,u,y,v)|) \}, 
\end{eqnarray*}
where $\mbox{Supp} (f(t))$ indicates the support of the function $(y,v) \to f(y,v,t)$. But for a continuum model, this maximum is likely to be infinite as soon as there exists an (even very small) non-zero density $f(y,v,t)$ for large values of the function $\Phi(|\dot \alpha(x,u,y,v)|$, $ |\tau(x,u,y,v)|)$. So, we replace the maximum by an average. We define:
\begin{eqnarray}
& & \hspace{-1cm} 
{\mathcal S}_\pm(x,u)  = \big\{ \, (y,v) \in {\mathbb R}^2 \times {\mathbb S}^1 \, \, \big| \, \,  \pm \dot \alpha(x,u,y,v) >0, \quad  \tau(x,u,y,v)>0, \nonumber \\
& & \hspace{3cm} 
D^2(x,u,y,v) < R^2, \quad |\dot \alpha(x,u,y,v)| < \sigma(|\tau(x,u,y,v)|) \, \big\}. 
\label{eq:S(v-u)}  
\end{eqnarray}
We note that, because of the last condition in (\ref{eq:S(v-u)}), we have $\Phi(|\dot \alpha|, |\tau|) >0$ on the set ${\mathcal S}_\pm(x,u)$. We then let $\Phi_\pm(x,u,t)$ the averages of $\Phi(|\dot \alpha|, |\tau|)$ over the set ${\mathcal S}_\pm(x,u)$, weighted by the distribution function of the pedestrians, namely: 
\begin{eqnarray}
& & \hspace{-1cm} 
\Phi_\pm(x,u,t)  = \frac{\int_{(y,v) \in {\mathcal S}_\pm(x,u), \, \eta \in {\mathbb R}^2} \,  \Phi(|\dot \alpha(x,u,y,v)|, |\tau(x,u,y,v)|) \, f(y,v,\eta,t) \, dy \, dv \, d\eta }
{\int_{(y,v) \in {\mathcal S}_\pm(x,u), \, \eta \in {\mathbb R}^2} f(y,v,\eta,t) \, dy \, dv \, d\eta }, \nonumber \\
& & \hspace{-1cm} 
\mbox{}
\label{eq:Phi+-} 
\end{eqnarray}
We can now define $\omega_f(x,u,\xi,t)$ by the following alternative: 

\begin{enumerate}
\item Small deviation to goal: If 
\begin{eqnarray}
& & \hspace{-1cm} 
- \Phi_{-} (x,u,t)\, \leq  \, \dot \alpha_{g} (x,u,\xi)  \, \leq  \, \Phi_{+} (x,u,t) . 
\label{eq:mf_deviation_small}
\end{eqnarray}
Then: 
\begin{eqnarray}
& & \hspace{-1cm} 
\omega_f (x,u,\xi,t) =  \nonumber \\
& & \hspace{-0.75cm} 
- \Phi_{+}(x,u,t) \, H \left( \big| \Phi_{-}(x,u,t) - |\dot \alpha_{g}(x,u,\xi)| \big| - \big| \Phi_{+}(x,u,t)-|\dot \alpha_{g}(x,u,\xi)| \big| \right) \nonumber \\
& & \hspace{-0.75cm} 
+ \Phi_{-}(x,u,t) \, H \left( \big| \Phi_{+}(x,u,t)-|\dot \alpha_{g}(x,u,\xi)| \big| - \big| \Phi_{-}(x,u,t) - |\dot \alpha_{g}(x,u,\xi)| \big| \right),
\label{eq:omegaf_case1}
\end{eqnarray}
where $H$ is again the Heaviside function. 
\item Large deviation to the goal:  
\begin{eqnarray}
& & \hspace{-1cm} 
\dot \alpha_{g}(x,u,\xi,t) \, <  \,  - \Phi_{-}(x,u,t) \quad \mbox{ or } \quad \Phi_{+}(x,u,t) <  \dot \alpha_{g}(x,u,\xi,t). 
\label{eq:mf_deviation_large}
\end{eqnarray}
Then, 
\begin{eqnarray}
& & \hspace{-1cm} 
\omega_f(x,u,\xi,t) = \dot \alpha_{g}(x,u,\xi).
\label{eq:omegaf_case2}
\end{eqnarray}
\end{enumerate}

Finally, the mean-field kinetic model consists of the kinetic equation (\ref{eq:kinetic}) for the distribution function $f(x,u,\xi,t)$, coupled with the mean-field force $F_f(x,u,\xi,t)$ given by (\ref{eq:mf_force}). The mean-field force is the elementary force acting on the particles located at $x$, with velocity $u$ and goal $\xi$ at time $t$. Its expression is given either by (\ref{eq:omegaf_case1}) or by (\ref{eq:omegaf_case2}) according to whether the corresponding particles have small or large deviation to the goal (respectively defined by the inequalities (\ref{eq:mf_deviation_small}) and (\ref{eq:mf_deviation_large})). This alternative depends on the mean-field evaluation of the average response towards pedestrians coming from the left or from the right given by (\ref{eq:Phi+-}). The expression of the force itself (\ref{eq:omegaf_case1}) depends on this response. The decision of making a left or right turn, reflected by the two Heaviside functions in  (\ref{eq:omegaf_case1}) is the one which minimizes the deviation to the goal. 

So far, we have taken the set of target point $\xi$ equal to the whole space ${\mathbb R}^2$. In practice, it is probably enough to deal with a finite number of target points. In this case, we would replace the continuous dependence of $f$ upon $\xi$ by a coupled system of a finite number of equations for $f_i(x,u,t)$, where $i=1,\ldots, I$ is the index of target points, and $I$ their total number. Most of the model equations would remain unchanged, except for (\ref{eq:Phi+-}) where the integrals over $\eta$ would be replaced by discrete summations. 

This description shows that the mean-field kinetic model is a direct statistical translation of the Individual-Based model described in section \ref{subsec_Nped}, up  to the transformation of the maximum operation in (\ref{eq:Phi+i}) and (\ref{eq:Phi-i}) into a mean-field average in (\ref{eq:Phi+-}). In section~\ref{sec:macro}, we use this kinetic model to derive several macroscopic models.

%%%%%%%%%%%%%%%%%%%%%%%%%%%%%%%%%%%%%%%%%%%%%%%%
%%%%%%%%%%%%%%%%%%%%%%%%%%%%%%%%%%%%%%%%%%%%%%%%
\subsection{Mean-field kinetic model for the potential-driven dynamics}
\label{sub:meanfield_modified}

Here, we investigate how the mean-field kinetic model of section \ref{sub:meanfield_deriv} must be adapted in the case of the potential-driven dynamics of section \ref{subsec_Nped_modif}. The only modification is the expression of the force (\ref{eq:mf_force}). Following section \ref{subsec_Nped_modif}, but using the averaging procedure of section \ref{sub:meanfield_deriv} instead of the 'max', we define the cost function $\Phi_c(x,u,t)$ for the cost of undergoing collisions with other pedestrians. We first introduce
\begin{eqnarray}
& & \hspace{-1cm} 
{\mathcal S}(x,u)  = \{ (y,v) \in {\mathbb R}^2 \times {\mathbb S}^1 \, \big| \,   \tau(x,u,y,v)>0, \quad D(x,u,y,v)^2 < R^2, \nonumber \\
& & \hspace{7cm}  
|\dot \alpha(x,u,y,v)| < \sigma(|\tau(x,u,y,v)|) \}. 
\label{eq:S(v-u)_modif}  
\end{eqnarray}
We then let $\Phi_c(x,u,t)$ be the average of $\Phi(|\dot \alpha|, |\tau|)$ over the set ${\mathcal S}(x,u)$, weighted by the distribution function of the pedestrians, namely: 
\begin{eqnarray}
& & \hspace{-1cm} 
\Phi_c(x,u,t)  = - \frac{\int_{(y,v) \in {\mathcal S}(x,u), \, \eta \in {\mathbb R}^2} \,  \Phi(|\dot \alpha(x,u,y,v)|, |\tau(x,u,y,v)|) \, f(y,v,\eta,t) \, dy \, dv \, d\eta }
{\int_{(y,v) \in {\mathcal S}(x,u), \, \eta \in {\mathbb R}^2} f(y,v,\eta,t) \, dy \, dv \, d\eta }, \nonumber \\
& & \hspace{-1cm} 
\mbox{}
\label{eq:Phi_c_mf} 
\end{eqnarray}
The cost of walking away from the target direction $\alpha_{g}$ is still given by (\ref{eq:Phi_t}). The total cost is then defined by
\begin{eqnarray}
& & \hspace{-1cm} 
\Phi(x,u,\xi,t) = \Phi_c(x,u,t) + \Phi_t(x,u,\xi). 
\label{eq:Phi_tot_mf} 
\end{eqnarray}
The force is obtained through: 
\begin{eqnarray}
& & \hspace{-1cm} 
F_f(x,u,\xi,t) = \omega_f(x,u,\xi,t)  u^\bot = - \nabla_u \Phi (x,u,\xi,t)  . 
\label{eq:Phi_grad_mf} 
\end{eqnarray}

Finally, the modified mean-field kinetic model for the potential-driven dynamics consists of eq.(\ref{eq:kinetic}), supplemented with the expression (\ref{eq:Phi_grad_mf}) of the force.

%%%%%%%%%%%%%%%%%%%%%%%%%%%%%%%%%%%%%%%%%%%%%%%%
%%%%%%%%%%%%%%%%%%%%%%%%%%%%%%%%%%%%%%%%%%%%%%%%
\subsection{Local approximations to the mean-field kinetic models}
\label{sub:meanfield_local}

We now propose spatially local approximations of the mean-field models for both the original and the potential-driven dynamics. We start with the original dynamics (section \ref{sub:meanfield_deriv}). 

\begin{figure}
\begin{center}
\input{Psi.pstex_t}
\caption{Graphical representation of formula (\ref{eq:Psi+-_local}). The domains $\hat {\mathcal S}_+(v-u)$ and $\hat {\mathcal S}_-(v-u)$ are represented by the blue and green shaded areas respectively (their left boundary is supposed to be outside the figure). The function $\hat \Phi$ depends only on $\widehat{\dot \alpha}$ and $\hat \tau$ which are constant along the dashed circles and dashed vertical lines respectively. The dashed circles are all circles centered along the vertical axis and passing through the origin. Indeed, we have $\widehat{\dot \alpha} = c \zeta_\bot |v-u|/\zeta^2$ and $c \tau = - \zeta_\parallel/|v-u|$. When integrating such a function on the shaded domains, the result obviously only depends on $|v-u|$.} 
\label{fig:Psi}
\end{center}
\end{figure}
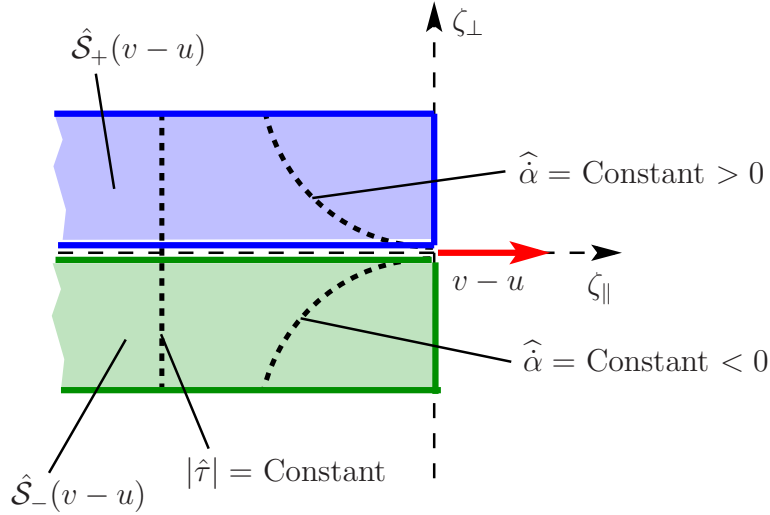

If we observe the system at a large distance, the various length scales involved in the interaction terms appear to be small. Therefore, under this assumption, it is legitimate to assume that there exists a small dimensionless quantity $\lambda \ll 1$ such that 
\begin{equation} 
R = \lambda \hat R, \quad \Phi_0 = \lambda \hat \Phi_0, \quad a = \frac{1}{\lambda} \hat a, \quad \tau_0 = \lambda \hat \tau_0, \quad b = \lambda^{c-1} \hat b, \quad \sigma = \frac{1}{\lambda} \hat \sigma
\label{eq:scaling_local_force_1}
\end{equation}
where all 'hat' quantities are assumed to be ${\mathcal O}(1)$. The scaling (\ref{eq:scaling_local_force_1}) is tailored to make the interaction force spatially local, while maintaining its temporal scale of order ${\mathcal O}(1)$, as we will see below. We introduce the change of variables $y = x + \lambda \zeta$, with $\zeta \in {\mathbb R}^2$ in all expressions involving $y$. We get the following expressions: 
\begin{eqnarray} 
&&\hspace{-1cm}
\dot \alpha (x,u,y,v) = \frac{1}{\lambda} \widehat{\dot \alpha} (\zeta, v-u) , \quad \widehat{\dot \alpha} (\zeta, v-u) = c \frac{\zeta \times (v-u)}{|\zeta|^2}, 
\label{eq:mf_alpha_local} \\
&&\hspace{-1cm}
\tau (x,u,y,v) = \lambda \hat \tau (\zeta,v-u), \quad \hat \tau (\zeta,v-u) = - \frac{1}{c} \frac{\zeta \cdot (v-u)}{|v-u|^2}, 
\label{eq:mf_tau_local} \\
&&\hspace{-1cm}
D (x,u,y,v) = \lambda \hat D(\zeta, v-u), \quad \hat D^2 (\zeta,v-u) = |\zeta|^2 - \Big( \zeta \cdot \frac{v-u}{|v-u|} \Big)^2. 
\label{eq:mf_Dmin_local} 
\end{eqnarray}
On the other hand, the DBA with respect to the goal $\xi$ is unchanged and still given by (\ref{eq:mf_alphag}). The function $\Phi(|\dot \alpha|, |\tau|)$ is changed into $\hat \Phi(|\widehat{\dot \alpha}|, |\hat \tau|)$, such that 
\begin{eqnarray} 
&&\hspace{-1cm}
\Phi(|\dot \alpha(x,y,u,v)|, |\tau(x,y,u,v)|) = \hat \Phi(|\widehat{\dot \alpha}(\zeta,v-u)|, |\hat \tau(\zeta,v-u)|), 
\label{eq:mf_Phi_local} \\
&&\hspace{-1cm}
\hat \Phi(|\widehat{\dot \alpha}|, |\hat \tau|) = \hat \Phi_0 \, \max \{ \hat \sigma(|\hat \tau|) - |\widehat{\dot \alpha}|, \, 0 \}, \quad \quad \hat \sigma(|\hat \tau|) = \hat a + \frac{\hat b}{(|\hat \tau| + \hat \tau_0)^c}, 
\label{eq:mf_sigma_local} 
\end{eqnarray}
Now, with this change of variables, formula (\ref{eq:Phi+-}) for $\Phi_+$ and $\Phi_-$ is written as follows: 
\begin{eqnarray}
& & \hspace{-1cm} 
\Phi_\pm(x,u,t)  = \frac{\int_{(\zeta, v, \eta) \in \breve {\mathcal S}_\pm(u)} \,  \Phi(|\widehat{\dot \alpha} (\zeta, v-u)|, |\hat \tau (\zeta, v-u)|) \, f(x + \lambda \zeta,v,\eta,t) \, d\zeta \, dv \, d\eta }
{\int_{(\zeta, v, \eta) \in \breve {\mathcal S}_\pm(u)} \,  f(x + \lambda \zeta,v,\eta,t) \, d\zeta \, dv \, d\eta  }, \nonumber \\
& & \hspace{-1cm} 
\mbox{}
\label{eq:Phi+-_chvar} 
\end{eqnarray}
where 
\begin{eqnarray*}
& & \hspace{-1cm}
\breve {\mathcal S}_\pm(u) = \{ (\zeta, v, \eta) \in {\mathbb R}^2 \times {\mathbb S}^1 \times {\mathbb R}^2 \, \,  | \, \, \pm \widehat{\dot \alpha} (\zeta, v-u) >0, \quad \hat \tau (\zeta, v-u) >0, \\
& & \hspace{2cm}
\hat D^2 (\zeta, v-u) < \hat R^2 (\zeta, v-u), \quad |\widehat{\dot \alpha}(\zeta, v-u)| < \sigma(|\hat \tau(\zeta, v-u)|)  \}. 
\end{eqnarray*}
Now, in the formal limit $\lambda \to 0$,  the dependence of $f$ upon $\zeta$ disappears and $\hat \Phi$ can be integrated out with respect to $\zeta$. Therefore,  formula (\ref{eq:Phi+-_chvar}) leads to: 
\begin{eqnarray}
& & \hspace{-1cm} 
\Phi_\pm(x,u,t)  = \frac{ \int_{(v,\eta) \in {\mathbb S}^1 \times {\mathbb R}^2}   \Psi_\pm(|v-u|) \, f(x,v,\eta,t) \, dv \, d\eta }
{\int_{(v,\eta) \in {\mathbb S}^1 \times {\mathbb R}^2}  \, f(x,v,\eta,t) \, dv \, d\eta }, 
\label{eq:Phi+-_local}  
\end{eqnarray}
where
\begin{eqnarray}
& & \hspace{-1cm} 
\Psi_\pm(|v-u|)  = \frac{1}{\mbox{Area}(\hat {\mathcal S}_\pm(v-u))} \int_{\zeta \in \hat {\mathcal S}_\pm(v-u)}  \hat \Phi(|\widehat{\dot \alpha}(\zeta, v-u)|, |\hat \tau(\zeta, v-u)|) \, d\zeta . 
\label{eq:Psi+-_local}  
\end{eqnarray}
We denote by $\hat {\mathcal S}_\pm(v-u)$ the set
\begin{eqnarray}
& & \hspace{-1cm} 
\hat {\mathcal S}_\pm(v-u)  = \{ \zeta \in {\mathbb R}^2 \, \big| \,  \pm \widehat{\dot \alpha}(\zeta, v-u) >0, \quad \hat \tau(\zeta, v-u)>0, \nonumber \\
& & \hspace{3cm} 
\hat D^2(\zeta, v-u) < \hat R^2, \quad |\widehat{\dot \alpha}(\zeta, v-u)| < \sigma(|\hat \tau(\zeta, v-u)|)  \},  
\label{eq:hatS(v-u)}  
\end{eqnarray}
and $\mbox{Area}(\hat {\mathcal S}_\pm(v-u))$ its two-dimensional area. It is a simple matter to check that $\hat {\mathcal S}_\pm(v-u) $ is a bounded domain as soon as $\hat a>0$ or $\hat a=0$ and $c<2$ (see (\ref{eq:sigma})), which we will suppose from now on. Therefore, $\mbox{Area}(\hat {\mathcal S}_\pm(v-u))$ is finite. The graphical representation of formula (\ref{eq:Psi+-_local}) can be found in Fig. \ref{fig:Psi}. The function $\Psi_\pm(|v-u|)$ can be computed numerically once for all. Once the functions $\Phi_\pm$ have been computed thanks to (\ref{eq:Phi+-_local}), the determination of $\omega_f$ follows the same procedure as in section \ref{sub:meanfield_deriv}, by means of eqs. (\ref{eq:mf_deviation_small}) through (\ref{eq:omegaf_case2}). 

In the case of the potential-driven dynamics (section \ref{sub:meanfield_modified}), the local approximation takes the following form. The formula (\ref{eq:Phi_c_mf}) for the cost function $\Phi_c$ associated to collisions with the other pedestrians is changed into:
\begin{eqnarray}
& & \hspace{-1cm} 
\Phi_c(x,u,t)  = - \frac{ \int_{(v,\eta) \in {\mathbb S}^1 \times {\mathbb R}^2}   \Psi(|v-u|) \, f(x,v,\eta,t) \, dv \, d\eta }
{\int_{(v,\eta) \in {\mathbb S}^1 \times {\mathbb R}^2}  \, f(x,v,\eta,t) \, dv \, d\eta }, 
\label{eq:Phi_local}  
\end{eqnarray}
where
\begin{eqnarray}
& & \hspace{-1cm} 
\Psi(|v-u|)  = \frac{1}{\mbox{Area}(\hat {\mathcal S}(v-u))} \int_{\zeta \in \hat {\mathcal S}(v-u)}  \hat \Phi(|\widehat{\dot \alpha}(\zeta,v-u)|, |\hat \tau(\zeta,v-u)|) \, d\zeta . 
\label{eq:Psi_local}  
\end{eqnarray}
and 
\begin{eqnarray}
& & \hspace{-1cm} 
\hat {\mathcal S}(v-u)  = \{ \zeta \in {\mathbb R}^2 \, \big| \,  \hat \tau(\zeta,v-u)>0, \quad \hat D(\zeta,v-u)^2 < \hat R^2, \nonumber \\
& & \hspace{7cm} 
|\widehat{\dot \alpha}(\zeta,v-u)| < \sigma(|\hat \tau(\zeta,v-u)|)  \}.  
\label{eq:hatS(v-u)_modif}  
\end{eqnarray}
Again, $\hat {\mathcal S}(v-u) $ is a bounded domain as soon as $\hat a>0$ or $\hat a=0$ and $c<2$. Formulas (\ref{eq:Phi_tot_mf}) and (\ref{eq:Phi_grad_mf}) remain unchanged. This local approximation of the potential-driven dynamics is the key of the hydrodynamic limit of section \ref{subsec_hydro}. 

The approximations developed in this section are called 'local' because the evaluation of the force $\omega_f$ at a point $x$ in space requires solely the knowledge of $f$ at the same point. Indeed, only the values of $f$ at point $x$ are needed to evaluate the integrals involved in (\ref{eq:Phi+-_local}) or in (\ref{eq:Phi_local}). The local approximation combined with the potential-driven dynamics as discussed in the previous paragraph are the two essential ingredients that allow for the hydrodynamic limit of section \ref{subsec_hydro}. 

%%%%%%%%%%%%%%%%%%%%%%%%%%%%%%%%%%%%%%%%%%%%%%%%
%%%%%%%%%%%%%%%%%%%%%%%%%%%%%%%%%%%%%%%%%%%%%%%%
%%%%%%%%%%%%%%%%%%%%%%%%%%%%%%%%%%%%%%%%%%%%%%%%
%%%%%%%%%%%%%%%%%%%%%%%%%%%%%%%%%%%%%%%%%%%%%%%%
\setcounter{equation}{0}
\section{Macroscopic models}
\label{sec:macro}

%%%%%%%%%%%%%%%%%%%%%%%%%%%%%%%%%%%%%%%%%%%%%%%%
%%%%%%%%%%%%%%%%%%%%%%%%%%%%%%%%%%%%%%%%%%%%%%%%
\subsection{Introduction to macroscopic models}
\label{subsec:macro_intro}

In this section, we develop macroscopic models for the mean-field kinetic model presented in the previous section. The development will closely follow \cite{Degond_etal_Heuristics} and we will only focus on the new computations, referring to \cite{Degond_etal_Heuristics} for the other ones.  
Macroscopic models are obtained by taking averages of functions of the particle velocity $u$ over the distribution function $f(x,u,\xi,t)$. The resulting macroscopic quantities are e.g. the density $\rho(x,\xi,t)$ or the mean velocity $U(x,\xi,t)$ of pedestrians at position $x$ with target point $\xi$ at time $t$:  
\begin{eqnarray}
& & \hspace{-1cm} 
\rho(x,\xi,t) = \int_{u \in {\mathbb S}^1} f(x,u,\xi,t) \, du, \label{eq:moments1} \\
& & \hspace{-1cm} 
U(x,\xi,t) = \frac{1}{\rho(x,\xi,t)} \, \int_{u \in {\mathbb S}^1} f(x,u,\xi,t) \, u \, du.
\label{eq:moments2}
\end{eqnarray}
Here, we keep the dependence of the macroscopic quantities over the target point $\xi$, as this information is very important in practical situations. It is also possible to introduce more global macroscopic quantities such as the total density $N(x,t)$ irrespective of their target point. In this case, $N$ would just be the integral of $\rho$ given by (\ref{eq:moments1}) over $\xi$. However, we will discard such models here. 

To pass from the kinetic model (\ref{eq:kinetic}) to a macroscopic model, one generally uses the moment method.  However, this method requires closure relations in order to terminate the hierarchy of moment equations. These are provided through an Ansatz which expresses $f$ as a function of $\rho$ and $U$. The justification of this Ansatz is sometimes possible through the so-called hydrodynamic limit, such as in gas dynamics (see e.g. \cite{Degond_review_Birkhauser_03}) and also  in the case of the heuristic-based pedestrian model of  \cite{Moussaid_etal_PNAS11} (see \cite{Degond_etal_Heuristics}).   

In this section, we propose three different closure hypotheses. The first one relies on a monokinetic distribution function and is valid in the strictly noiseless case. The second one, which can handle noisy cases, postulates that the distribution of velocities is a von Mises-Fisher (VMF) distribution in the velocity variable $u$. The VMF distribution is the most natural extension of the Gaussian to random variables belonging to the sphere \cite{Watson_JAP82}. Finally, the third one, which applies only to the potential-driven dynamics in the local approximation, relies on the hydrodynamic limit, in the same spirit as \cite{Degond_etal_Heuristics}.

%%%%%%%%%%%%%%%%%%%%%%%%%%%%%%%%%%%%%%%%%%%%%%%%
%%%%%%%%%%%%%%%%%%%%%%%%%%%%%%%%%%%%%%%%%%%%%%%%
\subsection{Monokinetic closure}
\label{subsec_monokinetic}

%%%%%%%%%%%%%%%%%%%%%%%%%%%%%%%%%%%%%%%%%%%%%%%%
\subsubsection{Monokinetic closure: derivation}
\label{subsub:monokinetic_derivation}

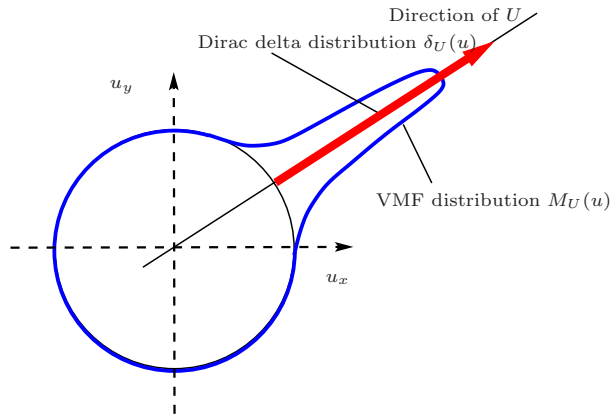
\begin{figure}[htbp]
\begin{center}
\input{VMF.pstex_t}
\caption{The Dirac delta distribution (in red) and the VMF distribution (in blue) as functions of $u$ in polar coordinates. The direction of the mean velocity $U$ is given by the black semi-line. The width of the VMF distribution about the velocity  $u=U/|U|$ at which it is maximal is a function of $\beta(|U|)$. In both cases, the mean velocity $U$ is a function of $(x,\xi,t)$ determined by the fluid model. We have $|U|=1$ in the Dirac delta distribution case and $|U|<1$ in the VMF case. }
\label{fig:VMF}
\end{center}
\end{figure}

In this section, we consider the mean-field kinetic model (\ref{eq:kinetic}) without noise. The equation is written: 
\begin{eqnarray}
& & \hspace{-1cm} \partial_t f + c u \cdot \nabla_x f + \nabla_u \cdot ( F_f \, f) = 0 , \label{mf_f_nn}
\end{eqnarray}
and is coupled to (\ref{eq:mf_force}) with $\omega_f$ given by (\ref{eq:omegaf_case1}) or (\ref{eq:omegaf_case2}). The monokinetic Ansatz is written:
\begin{eqnarray}
& & \hspace{-1cm} f(x,u,\xi,t) = \rho(x,\xi,t) \delta_{U(x,\xi,t)}(u), \label{mk_closure}
\end{eqnarray}
where $\delta_{U}(u)$ is the Dirac delta located at $U$. Note that, by definition, $U(x,\xi,t) \in {\mathbb S}^1$ i.e. is a vector of norm $1$. The monokinetic distribution function is graphically represented in Fig. \ref{fig:VMF}. Easy computations \cite{Degond_review_Birkhauser_03} show that (\ref{mk_closure}) is an exact solution to (\ref{mf_f_nn}) provided $\rho$ and $U$ satisfy: 
\begin{eqnarray}
& & \hspace{-1cm} \partial_t \rho + \nabla_x \cdot (c \rho U) = 0 , \label{mk_rho} \\
& & \hspace{-1cm} \partial_t U + c U \cdot \nabla_x  U = \tilde F(x,\xi,t) , \label{mk_U} 
\end{eqnarray}
with 
\begin{eqnarray}
\tilde F(x,\xi,t) = \tilde \omega_{\rho,U} (x,\xi,t) \, U^\bot(x,\xi,t), \label{mk_F} 
\end{eqnarray}
where
\begin{eqnarray}
\tilde \omega_{\rho,U}(x,\xi,t) := \omega_{\rho \delta_{U}} (x,U(x,\xi,t),\xi,t) , 
\label{mk_omega} \end{eqnarray}
and where $\omega_f$ with $f = \rho \delta_{U}$ is given by the procedure detailed in section \ref{sub:meanfield_deriv}. Specifically, letting $u=U(x,\xi,t)$ and $v= U(y,\eta,t)$ in all formulas (\ref{eq:mf_alpha}) to (\ref{eq:mf_alphag}) defines functions $\widetilde{\dot \alpha} (x,\xi,y,\eta,t)$, $\tilde \tau (x,\xi,y,\eta,t)$, $\tilde D (x,\xi,y,\eta,t)$, $\widetilde{\dot \alpha_g} (x,\xi,t)$. Then, we define 
\begin{eqnarray}
& & \hspace{-1cm} 
\tilde {\mathcal S}_\pm(x,\xi,t)  = \big\{ \, (y,\eta) \in {\mathbb R}^2 \times {\mathbb R}^2 \, \big| \,  \pm \widetilde{\dot \alpha}(x,\xi,y,\eta,t) >0, \quad  \tilde \tau (x,\xi,y,\eta,t) >0, \nonumber \\
& & \hspace{2cm} 
\tilde D(x,\xi,y,\eta,t)^2 < R^2, \quad |\widetilde{\dot \alpha}(x,\xi,y,\eta,t)| < \sigma(|\tilde \tau(x,\xi,y,\eta,t)|) \, \big\} , 
\label{eq:tilde_S}  
\end{eqnarray}
and
\begin{eqnarray}
& & \hspace{-1cm} 
\tilde \Phi_\pm(x,\xi,t)  = \frac{\int_{(y,\eta) \in \tilde {\mathcal S}_\pm(x,\xi,t)}   \Phi(|\widetilde{\dot \alpha}(x,\xi,y,\eta,t)|, |\tilde \tau(x,\xi,y,\eta,t)|) \, \rho(y,\eta,t) \, dy \, d\eta }
{\int_{(y,\eta) \in \tilde {\mathcal S}_\pm(x,\xi,t)}  \rho(y,\eta,t) \, dy \, d\eta }. 
\label{eq:macro_Phi+} 
\end{eqnarray}
Then, $\tilde \omega_{\rho,U}$ is given by formulas (\ref{eq:mf_deviation_small}) to (\ref{eq:omegaf_case2}), putting tildes on all quantities. 

The macroscopic model with monokinetic closure is obtained by collecting (\ref{mk_rho}), (\ref{mk_U}), (\ref{mk_F}). Eq. (\ref{mk_rho}) is the continuity equation for the mass while (\ref{mk_U}) expresses how the fluid velocity evolves as a consequence of the pedestrian interactions. The force term~(\ref{mk_F}) describes the rotation of the pedestrian in response to close encounters. By the definition~(\ref{mk_F}), $\tilde F \cdot U = 0$. Then, it follows that the constraint $|U|=1$ is satisfied at any time provided it is satisfied at initial time \cite{Degond_etal_Heuristics}. The intensity of the force is non-local in space. This non-locality is due to the fact that the pedestrian anticipates the behavior of neighboring pedestrians to take a decision. 

%%%%%%%%%%%%%%%%%%%%%%%%%%%%%%%%%%%%%%%%%%%%%%%%
\subsubsection{Monokinetic closure for the potential-driven dynamics}
\label{subsub:monokinetic_modified}

If instead of the original dynamics, the mean-field model for the potential-driven dynamics is used, the computation of $\tilde F$ follows the procedure described from eq. (\ref{eq:S(v-u)_modif}) through (\ref{eq:Phi_grad_mf}). Specifically, we define 
\begin{eqnarray}
& & \hspace{-1.5cm} 
\tilde {\mathcal S}(x,u,t)  = \big\{ \, (y,\eta) \in {\mathbb R}^2 \times {\mathbb R}^2 \, \big| \,  \tau (x,u,y,U(y,\eta,t)) >0, \nonumber \\
& & \hspace{-1.0cm} 
D(x,u,y,U(y,\eta,t))^2 < R^2, \quad |\dot \alpha(x,u,y,U(y,\eta,t))| < \sigma(|\tau(x,u,y,U(y,\eta,t))|) \, \big\}. 
\label{eq:tilde_S_modif}  
\end{eqnarray}
Then, the cost function associated to the collisions with the other pedestrians is given by:
\begin{eqnarray}
& & \hspace{-1cm} 
\Phi_c(x,u,t)  = \nonumber \\
& & \hspace{-1cm} 
= - \frac{\int_{(y,\eta) \in \tilde {\mathcal S}(x,u,t)}   \Phi(|\dot \alpha(x,u,y,U(y,\eta,t))|, |\tau(x,u,y,U(y,\eta,t))|) \, \rho(y,\eta,t) \, dy \, d\eta }
{\int_{(y,\eta) \in \tilde {\mathcal S}(x,u,t)}  \rho(y,\eta,t) \, dy \, d\eta }. 
\label{eq:macro_Phi_mk_modif} 
\end{eqnarray}
Finally, the force term is given by
\begin{eqnarray}
& & \hspace{-1cm} 
F_f(x,\xi,t) = - \nabla_u \Phi (x,U(x,\xi,t),\xi,t)  . 
\label{eq:Phi_grad_mk} 
\end{eqnarray}
with the total cost $\Phi$ given by (\ref{eq:Phi_tot_mf}) as the sum of the cost associated to collisions (\ref{eq:macro_Phi_mk_modif}) and of the target cost given by (\ref{eq:Phi_t}).

%%%%%%%%%%%%%%%%%%%%%%%%%%%%%%%%%%%%%%%%%%%%%%%%
\subsubsection{Monokinetic closures: local approximations}
\label{subsub:monokinetic_local}

Local approximations of the monokinetic closures of both the original and potential-driven dynamics can be given, following section \ref{sub:meanfield_local}. We first consider the monokinetic closure of the original model (section \ref{subsub:monokinetic_derivation}). Applying (\ref{eq:Phi+-_local}) with $f$ given by (\ref{mk_closure}) and evaluating it for $u = U(x,\xi,t)$, the spatially local approximation of $\tilde \Phi_\pm$ can be derived and leads to:
\begin{eqnarray}
& & \hspace{-1cm} 
\tilde \Phi_\pm(x,\xi,t)  = \frac{ \int_{\eta \in {\mathbb R}^2}   \Psi_\pm(|U(x,\eta,t)-U(x,\xi,t)|) \, \rho(x,\eta,t) \, d\eta }
{\int_{\eta \in {\mathbb R}^2}  \rho(x,\eta,t)\, d\eta }, 
\label{eq:tildePhi+-_local}  
\end{eqnarray}
with $\Psi_\pm$ given by (\ref{eq:Psi+-_local}). Once the functions $\tilde \Phi_\pm$ have been computed thanks to (\ref{eq:Phi+-_local}), the determination of $\tilde \omega_{\rho,U}$ follows the same procedure as in section \ref{sub:meanfield_deriv}. 

In the case where the potential-driven dynamics is combined with the local approximation of section \ref{subsub:monokinetic_local}, the cost function associated to the collisions with the other pedestrians takes the form
\begin{eqnarray}
& & \hspace{-1cm} 
\Phi_c(x,u,t)  = - \frac{ \int_{\eta \in {\mathbb R}^2}   \Psi(|U(x,\eta,t)-u)|) \, \rho(x,\eta,t) \, d\eta }
{\int_{\eta \in {\mathbb R}^2}  \rho(x,\eta,t)\, d\eta }, 
\label{eq:tildePhi+-_local_modif}  
\end{eqnarray}
with $\Psi$ given by (\ref{eq:Psi_local}). Then, the force $F_f$ is given by (\ref{eq:Phi_grad_mk}) with the total cost $\Phi$ given by (\ref{eq:Phi_tot_mf}) and the target cost given by (\ref{eq:Phi_t}). 

Like the monokinetic fluid model of \cite{Degond_etal_Heuristics},  these models, be they nonlocal or local, are of pressureless gas dynamics type \cite{Bouchut_AdvKinTheoryComputing94}. Indeed, the left-hand side of (\ref{mk_U}) describes the convection of a fluid with zero pressure. The force term at the right-hand side of (\ref{mk_U}) does not contain any spatial gradients. Therefore, large density gradients due to the formation of shock waves in (\ref{mk_U}) cannot be counterbalanced by the force term. In this case, mass concentrations may be produced and the model breaks down once these concentrations appear. The VMF closure developed in the next section attempts at providing a cure to this deficiency. Indeed, pressure is associated to the kinetic velocities of the particles, i.e. their deviation to the  local mean velocity. In the monokinetic closure, this deviation is zero and consequently there is no pressure. The VMF closure is associated to non-zero kinetic velocities and is likely to restore the influence of the pressure.

%%%%%%%%%%%%%%%%%%%%%%%%%%%%%%%%%%%%%%%%%%%%%%%%
%%%%%%%%%%%%%%%%%%%%%%%%%%%%%%%%%%%%%%%%%%%%%%%%
\subsection{VMF closure}
\label{subsec:gaussian}

%%%%%%%%%%%%%%%%%%%%%%%%%%%%%%%%%%%%%%%%%%%%%%%%
\subsubsection{VMF closure: derivation}
\label{subsub:VMF_derivation}

We first derive the moment equations of the noisy mean-field kinetic equation (\ref{eq:kinetic}). Integrating (\ref{eq:kinetic}) with respect to $u$ against the functions $1$ and $u$ leads to the mass and momentum balance equations. The algebra is the same as in \cite{Degond_etal_Heuristics} and we refer the reader to it for details. We get: 
\begin{eqnarray}
& & \hspace{-1cm} \partial_t \rho + \nabla_x \cdot (c \rho U) = 0 , \label{gauss_rho}  \\
& & \hspace{-1cm} 
\partial_t (\rho U) + \nabla_x \cdot ( c S ) = \rho {\mathcal F} - d \rho U  , \label{gauss_U}  
\end{eqnarray}
with the $2 \times 2$ tensor $S$ defined by
$$ S = \int_{u \in {\mathbb S}^1} f \, u \otimes u \, du, $$
and $u \otimes u$ is a matrix of components $(u \otimes u)_{ij} = u_i \, u_j$. The macroscopic force ${\mathcal F}$ is given by: 
\begin{eqnarray}
{\mathcal F} (x,\xi,t) &=& \rho^{-1}(x, \xi, t)   \, \int_{u \in {\mathbb S}^1}   F_f(x,u,\xi,t) \,  f(x,u,\xi, t)  \, du \nonumber \\ 
&=& \rho^{-1}(x, \xi, t)   \, \int_{u \in {\mathbb S}^1}   \omega_f(x,u,\xi,t) \, u^\bot \,  f(x,u,\xi, t)  \, du.
\label{eq:gauss_F}
\end{eqnarray}

To express $S$ and ${\mathcal F}$ analytically, we need a closure assumption, i.e. a prescription for~$f$. Like in \cite{Degond_etal_Heuristics}, we assume that $f$ is a von Mises-Fisher (VMF) distribution about the mean direction $U$. The VMF distribution is discussed in \cite{Watson_JAP82}. It is given by 
\begin{eqnarray}
& & \hspace{-1cm} 
M_U(u) = \frac{1}{Z} \exp\{ \beta \, (u \cdot \Omega) \}, \quad \Omega = \frac{U}{|U|}, 
\label{eq:VMF_norm}
\end{eqnarray}
where $\beta$ plays the role of an inverse temperature (we will see that  $\beta$ is related to $|U|$). The quantity $Z$ is a normalizing constant such that $M_U$ is a probability density on ${\mathbb S}^1$. It does not depend on $\Omega$ and is given by: 
\begin{eqnarray}
Z =\ Z(\beta) = 2 \pi I_0(\beta) .
\label{eq:Z}
\end{eqnarray}
We recall that  $I_k(x)$ denotes the modified Bessel function of the first kind:
$$ I_k(x) = \frac{1}{\pi} \int_0^{\pi} \exp\{ x \, \cos \theta \} \, \cos (k \, \theta) \, d \theta, \quad \forall x \in {\mathbb R}, \quad  \forall k \in {\mathbb N}.  $$
The VMF  distribution function is graphically represented in Fig. \ref{fig:VMF}. 
We note that the flux of $M_U$ is given by:
\begin{eqnarray}
& & \hspace{-1cm} 
\int_{u \in {\mathbb S}^1} M_U(u) \, u \, du = \int_{u \in {\mathbb S}^1} M_U(u) \, (u \cdot \Omega) \, du \,\, \Omega = \frac{I_1(\beta)}{I_0(\beta)} \, \Omega. 
\label{eq:currVMF}
\end{eqnarray}
The VMF Ansatz is written:
\begin{eqnarray}
& & \hspace{-1cm} 
f(x,u,\xi,t) = \rho(x,\xi,t) M_{U(x,\xi,t)} (u), 
\label{eq:VMF}
\end{eqnarray}
where $\rho(x,\xi,t)$ and $U(x,\xi,t)$ are the moments (\ref{eq:moments1}) and (\ref{eq:moments2}) of $f$. In view of (\ref{eq:currVMF}), the consistency with (\ref{eq:moments2}) requires that $\beta$ and $|U|$ be linked by the relation:
\begin{eqnarray}
\frac{I_1(\beta)}{I_0(\beta)} =  |U|.
\label{eq:compa2}
\end{eqnarray}
This equation has a unique solution $\beta(|U|) \in [0,\infty)$ for all $|U|$ such that $|U|<1$ (see \cite{Degond_etal_arXiv:1109.2404}). The condition $|U|<1$ is consistent with the kinetic model. Indeed, the microscopic velocities $u$  satisfy $|u|=1$. Therefore, the parameter $|U|$ acts as an order parameter. When $|U|$ is close to zero, the VMF distribution is almost isotropic while when $|U|$ is close to $1$, the VMF distribution is like a Dirac delta at the velocity direction $\Omega$  (see e.g.  \cite{Degond_etal_arXiv:1109.2404, Gregoire_Chate_PRL04, Vicsek_etal_PRL95} for the role of the order parameter in self-propelled particle systems). We will now write $\beta = \beta (|U|)$, $Z = Z(|U|)$, which leads to the following expression of the VMF Ansatz (omitting the dependences of $\rho$ and $U$ upon $(x,\xi,t)$  for clarity):
\begin{eqnarray}
& & \hspace{-1cm} 
f(u) = \rho M_U(u) = \rho \frac{1}{Z(|U|)} \exp\left\{ \frac{\beta(|U|)}{|U|} \, (u \cdot U) \right\}.  
\label{eq:VMF2}
\end{eqnarray}

Now, with (\ref{eq:VMF2}), the tensor $S$ can be computed \cite{Degond_etal_Heuristics}, and is given by: 
\begin{eqnarray}
&&\hspace{-1cm}
S = \rho \, \big(\gamma_\parallel(|U|) \, U \otimes U \, + \,  \gamma_\bot(|U|) \, U^\bot \otimes U^\bot \big) , 
\label{eq:S}
\end{eqnarray}
with 
\begin{eqnarray}
&&\hspace{-1cm}
\gamma_\parallel(|U|) = \frac{1}{2 |U|^2} (1 + \frac{I_2(\beta)}{I_0(\beta)}), \quad  \gamma_\bot(|U|) = \frac{1}{2 |U|^2} (1 - \frac{I_2(\beta)}{I_0(\beta)}), 
\label{eq:gamma}
\end{eqnarray}
and $\beta = \beta(|U|)$. Since $I_2/I_0 < 1$, the matrix $S$ is positive definite. In the limit $\beta \to \infty$, $ S \to \rho U \otimes U$, and we recover the expression of the monokinetic closure (second term at the left-hand side of (\ref{mk_U})). 

We now consider the force term (\ref{eq:gauss_F}). Using (\ref{eq:VMF}), we have:
\begin{eqnarray}
{\mathcal F} (x,\xi,t) &=&  \, \int_{u \in {\mathbb S}^1}   F_{\rho M_U}(x,u,\xi,t)  \,  M_{U(x,\xi,t)}(u)  \, du \nonumber \\
&=& \int_{u \in {\mathbb S}^1}   \omega_{\rho M_U}(x,u,\xi,t) \, u^\bot \,  M_{U(x,\xi,t)}(u)  \, du, 
\label{eq:gauss_F_2}
\end{eqnarray}
where $\omega_f$ with $f = \rho M_U$ is given by the procedure detailed in section \ref{sub:meanfield_deriv}. Taking advantage of the VMF Ansatz (\ref{eq:VMF}), we can write:
\begin{eqnarray}
& & \hspace{-1cm} 
\Phi_\pm(x,u,t)  = \frac{ \int_{(y,\eta) \in {\mathbb R}^2 \times {\mathbb R}^2} {\mathcal H}_\pm (x,u,y,U(y,\eta,t)) \, \rho(y,\eta,t) \, dy \, d\eta }
{\int_{(y,\eta) \in {\mathbb R}^2 \times {\mathbb R}^2} {\mathcal H}_{0\pm} (x,u,y,U(y,\eta,t)) \, \rho(y,\eta,t) \, dy \, d\eta }, 
\label{eq:barPhi+} 
\end{eqnarray}
with
\begin{eqnarray}
& & \hspace{-1cm} 
{\mathcal H}_\pm (x,u,y,U)  = \int_{v \in \Sigma_\pm(x,u,y)}  \Phi(|\dot \alpha(x,u,y,v)|, |\tau(x,u,y,v)|) \, M_{U}(v) \, dv, 
\label{eq:calH+} \\
& & \hspace{-1cm} 
{\mathcal H}_{0\pm} (x,u,y,U)  = \int_{v \in \Sigma_\pm(x,u,y)}  M_{U}(v) \, dv, 
\label{eq:calH0+} 
\end{eqnarray}
where the set $\Sigma_\pm(x,u,y)$ is defined by:
\begin{eqnarray}
& & \hspace{-1cm} 
\Sigma_\pm (x,u,y)  = \{ v \in {\mathbb S}^1 \, \big| \,  \pm \dot \alpha(x,u,y,v) >0, \quad  \tau(x,u,y,v)>0, \nonumber \\
& & \hspace{2cm} 
\quad D(x,u,y,v)^2 < R^2, \quad |\dot \alpha(x,u,y,v)| < \sigma(|\tau(x,u,y,v)|) \}. 
\label{eq:Sigma}  
\end{eqnarray}
Then, $\omega_{\rho M_U}(x,u,\xi,t)$ is given by formulas (\ref{eq:mf_deviation_small}) to (\ref{eq:omegaf_case2}), exactly like in section \ref{sub:meanfield_deriv}. The functions ${\mathcal H}_\pm$ and ${\mathcal H}_{0\pm}$ can be computed numerically a priori.

We now summarize the macroscopic model. It consists of the system: 
\begin{eqnarray}
& & \hspace{-1cm} 
\partial_t \rho + \nabla_x \cdot (c \rho U) = 0 . 
\label{gauss_rho_2} \\
& & \hspace{-1cm} 
\partial_t (\rho U) + \nabla_x \cdot \big( c \rho  \, (\gamma_\parallel \, U \otimes U \, + \,  \gamma_\bot \, U^\bot \otimes U^\bot ) \big) =  \rho \, {\mathcal F} - d \rho U, \label{gauss_U_2}  
\end{eqnarray}
coupled with (\ref{eq:gauss_F_2}). This a system for $\rho(x,\xi,t)$ and $U(x,\xi,t)$. Like the monokinetic closure, it is composed of the continuity equation for the mass density (\ref{gauss_rho_2}) and a balance equation for the fluid momentum (\ref{gauss_U_2}).  By contrast with the monokinetic closure, the left-hand side of the momentum eq. (\ref{gauss_U_2}) is expressed in divergence form. Compared to standard fluid-dynamic models, the transport operator has an unusual form, with the occurrence of the tensor $U^\bot \otimes U^\bot$, which is nothing but the adjugate of matrix $U \otimes U$ (i.e. the transpose of its cofactor matrix). This term occurs as a consequence of the non-classical closure using the VMF distribution. 

The target point $\xi$ appears implicitly through the force. The bulk force acting on a fluid element, given through expression (\ref{eq:gauss_F_2}) consists of an average of the elementary force $\omega_{M_U}$ over the VMF distribution. The computation of the elementary force itself involves the VMF distribution (hence the notation $\omega_{M_U}$) through the computation of the collision indicators $\Phi_\pm$ (see (\ref{eq:barPhi+})). Due to the known dependence of $M_U(v)$ on $v$ the quantities ${\mathcal H}_\pm$ and ${\mathcal H}_{0\pm}$ can be precalculated through (\ref{eq:calH+}), (\ref{eq:calH0+}). The resulting expression of the force is non-local and translates the anticipation capacity of the pedestrians.

%%%%%%%%%%%%%%%%%%%%%%%%%%%%%%%%%%%%%%%%%%%%%%%%
\subsubsection{VMF closure for the potential-driven dynamics}
\label{subsub:VMF_modified}

In this section, we adapt the previous VMF closure to the potential-driven dynamics of section \ref{sub:meanfield_modified}. The only quantity that changes is the fluid force ${\mathcal F}$ given by (\ref{eq:gauss_F_2}). We introduce the cost of collisions with the other pedestrians:  
\begin{eqnarray}
& & \hspace{-1cm} 
\Phi_c(x,u,t)  = - \frac{ \int_{(y,\eta) \in {\mathbb R}^2 \times {\mathbb R}^2} {\mathcal H} (x,u,y,U(y,\eta,t)) \, \rho(y,\eta,t) \, dy \, d\eta }
{\int_{(y,\eta) \in {\mathbb R}^2 \times {\mathbb R}^2} {\mathcal H}_{0} (x,u,y,U(y,\eta,t)) \, \rho(y,\eta,t) \, dy \, d\eta }, 
\label{eq:barPhi_modif} 
\end{eqnarray}
with
\begin{eqnarray}
& & \hspace{-1cm} 
{\mathcal H} (x,u,y,U)  = \int_{v \in \Sigma(x,u,y)}  \Phi(|\dot \alpha(x,u,y,v)|, |\tau(x,u,y,v)|) \, M_{U}(v) \, dv, 
\label{eq:calH_modif} \\
& & \hspace{-1cm} 
{\mathcal H}_{0} (x,u,y,U)  = \int_{v \in \Sigma(x,u,y)}  M_{U}(v) \, dv, 
\label{eq:calH0_modif} 
\end{eqnarray}
where the set $\Sigma(x,u,y)$ is defined by:
\begin{eqnarray}
& & \hspace{-1cm} 
\Sigma (x,u,y)  = \{ v \in {\mathbb S}^1 \, \big| \,  \tau(x,u,y,v)>0, \nonumber \\
& & \hspace{2cm} 
\quad D(x,u,y,v)^2 < R^2, \quad |\dot \alpha(x,u,y,v)| < \sigma(|\tau(x,u,y,v)|) \}. 
\label{eq:Sigma_modif}  
\end{eqnarray}
Then, we let 
\begin{eqnarray}
& & \hspace{-1cm} 
F_{\rho M_U}(x,u,\xi,t) = - \nabla_u \Phi (x,u,\xi,t)  , 
\label{eq:Phi_grad_vmf} 
\end{eqnarray}
where $\Phi$ is the total cost (\ref{eq:Phi_tot_mf}) given by the sum of the collision cost $\Phi_c$ (\ref{eq:barPhi_modif}) and the target cost $\Phi_t$ (\ref{eq:Phi_t}). Using the stokes formula, we can take advantage of the gradient form of (\ref{eq:Phi_grad_vmf}) to simplify expression (\ref{eq:gauss_F_2}) and get:
\begin{eqnarray}
&& \hspace{-1cm}
{\mathcal F} (x,\xi,t) = \beta \big(|U(x,\xi,t)| \big)  \int_{u \in {\mathbb S}^1}  \Phi(x,u,t)  \,  M_{U(x,\xi,t)}(u) \, du \, \Omega(x,\xi,t)  \nonumber \\
&& \hspace{-0.6cm} 
-   \int_{u \in {\mathbb S}^1}  \Phi(x,u,t)  \,  M_{U(x,\xi,t)}(u) \, \Big(1 + \beta \big(|U(x,\xi,t)| \big) \big( u \cdot \Omega(x,\xi,t) \big)\Big) \, u \, du , 
\label{eq:gauss_F_local_2}
\end{eqnarray}
with $\Omega(x,\xi,t) = U(x,\xi,t)/|U(x,\xi,t)|$. The computation is detailed in \cite{Degond_etal_Heuristics}.

%%%%%%%%%%%%%%%%%%%%%%%%%%%%%%%%%%%%%%%%%%%%%%%%
\subsubsection{VMF closures: local approximations}
\label{subsub:VMF_local}

Like for the monokinetic closure case, local approximations can be given. For the original mean-field model (section \ref{subsub:VMF_derivation}), applying (\ref{eq:Phi+-_local}) with $f$ given by (\ref{eq:VMF}) leads to:
\begin{eqnarray}
& & \hspace{-1cm} 
\Phi_\pm(x,u,t)  = \frac{ \int_{\eta \in {\mathbb R}^2}   \tilde \Psi_\pm(u,U(x,\eta,t)) \, \rho(x,\eta,t) \, d\eta }
{\int_{\eta \in {\mathbb R}^2}  \, \rho(x,\eta,t)  \, d\eta }, 
\label{eq:VMF_Phi+-_local}  
\end{eqnarray}
with
\begin{eqnarray}
& & \hspace{-1cm} 
\tilde \Psi_\pm(u,U)  = \int_{v \in {\mathbb S}^1 }   \Psi_\pm(|v-u|) \, M_{U}(v) \, dv , 
\label{eq:VMF_tilde_Phi+-_local}  
\end{eqnarray}
and with $\Psi_\pm$ given by (\ref{eq:Psi+-_local}). The determination of $\omega_{\rho M_U}$ then follows the same procedure as in section \ref{sub:meanfield_deriv}.

For the potential-driven dynamics (section \ref{subsub:VMF_modified}), the cost of collisions with the other pedestrians in the local approximation is given by:  
\begin{eqnarray}
& & \hspace{-1cm} 
\Phi(x,u,t)  = \frac{ \int_{\eta \in {\mathbb R}^2}   \tilde \Psi(u,U(x,\eta,t)) \, \rho(x,\eta,t) \, d\eta }
{\int_{\eta \in {\mathbb R}^2}  \, \rho(x,\eta,t)  \, d\eta }, 
\label{eq:VMF_Phi_local_modif}  
\end{eqnarray}
with
\begin{eqnarray}
& & \hspace{-1cm} 
\tilde \Psi(u,U)  = \int_{v \in {\mathbb S}^1 }   \Psi(|v-u|) \, M_{U}(v) \, dv , 
\label{eq:VMF_tilde_Phi_local_modif}  
\end{eqnarray}
and with $\Psi$ given by (\ref{eq:Psi_local}). Then, the force ${\mathcal F}$ is given by (\ref{eq:gauss_F_local_2}). 

The VMF closure of the potential-driven dynamics in the local approximation yields the simplest fluid model. The force potential (\ref{eq:VMF_Phi_local_modif}) is obtained through a local average of the function $\tilde \Psi$ over the density of particles having given target points $\eta$. The function $\tilde \Psi$ itself is some kind of measure of the distance between the velocity $u$ (at which the potential is evaluated) and the local average fluid velocity $U(x,\eta,t)$. Once the force potential is known, it can be averaged over the VMF distribution in order to get an estimate of the fluid force (formula (\ref{eq:gauss_F_local_2})).

%%%%%%%%%%%%%%%%%%%%%%%%%%%%%%%%%%%%%%%%%%%%%%%%
%%%%%%%%%%%%%%%%%%%%%%%%%%%%%%%%%%%%%%%%%%%%%%%%
\subsection{Hydrodynamic limit of the potential-driven dynamics in the local approximation}
\label{subsec_hydro}

In this section, we focus on the potential-driven dynamics in its local approximation and we discuss the hydrodynamic limit of the associated mean-field kinetic model (described in the second part of section \ref{sub:meanfield_local}). In the hydrodynamic limit, the interaction force $F_f$ and the noise diffusion constant $d$ are very large i.e. there exists a small parameter $\varepsilon \ll 1$ such that both $F_f$ and $d$ can be rescaled as follows 
\begin{equation} 
F_f = \frac{1}{\varepsilon} \hat F_f, \qquad d = \frac{1}{\varepsilon} \hat d 
\label{eq:scaling_hydro}
\end{equation}
Under this scaling, the mean-field model (\ref{eq:kinetic}) is written (omitting the 'hats' for simplicity): 
\begin{eqnarray}
&& \hspace{-1cm} 
\partial_t f^\varepsilon + c u \cdot \nabla_x f^\varepsilon = \frac{1}{\varepsilon} Q_{\Phi_{f^\varepsilon}}(f^\varepsilon). 
\label{eq:hydro_f}
\end{eqnarray}
Here, the collision operator $Q_{\Phi_f}(f)$ collects the negative of the last term of the left-hand side of (\ref{eq:kinetic}) (modeling the reaction of the pedestrians to the collisions with the other pedestrians) and the diffusion term at the right-hand side of (\ref{eq:kinetic})  (which models the noise). We have parametrized the collision operator by the potential $\Phi_f$ and highlighted the dependence of the potential on $f$ (through (\ref{eq:Phi_local})). In the local approximation, the collision operator $Q_{\Phi_f}(f)$ operates only on $u$ and $\xi$, leaving $(x,t)$ as mere parameters. Therefore, we consider it as acting on functions of $f(u,\xi)$ only. For a given function $(u,\xi) \in {\mathbb S}^1 \times {\mathbb R}^2 \to \Phi(u,\xi) \in {\mathbb R}$, the expression of the collision operator is:
\begin{eqnarray}
&& \hspace{-1cm} 
Q_\Phi(f) = -  \nabla_u \cdot ( - \nabla_u \Phi \, f) + d \Delta_u f . 
\label{eq:hydro_Q}
\end{eqnarray}
For a given function $f(u,\xi)$, $\Phi_f(u,\xi)$ is defined by 
\begin{eqnarray}
&& \hspace{-1cm} \Phi_f(u,\xi) = \Phi_t(u,\xi) - \frac{ \int_{(v,\eta) \in {\mathbb S}^1 \times {\mathbb R}^2}   \Psi(|v-u|) \, f(v,\eta) \, dv \, d\eta }
{\int_{(v,\eta) \in {\mathbb S}^1 \times {\mathbb R}^2}  \, f(v,\eta) \, dv \, d\eta }, 
\label{eq:hydro_Phi_local}
\end{eqnarray}
where $\Phi_t$ and $\Psi (|v-u|)$ are the known functions given by (\ref{eq:Phi_t}) and (\ref{eq:Psi_local}). In this formula, we have omitted the dependences of $\Phi_t$ on $x$ and of $f$ on $(x,t)$, because they are mere parameters. 

If the limit $\varepsilon \to 0$ is formally taken in (\ref{eq:hydro_f}) and if we assume that there exists a smooth function $f^0$ such that 
\begin{eqnarray}
&& \hspace{-1cm} 
f^\varepsilon \longrightarrow f^0 \quad \mbox{ as }  \quad \varepsilon \longrightarrow 0, 
\label{eq:convergence}
\end{eqnarray}
smoothly, then, we find that $f^0$ is necessarily a solution of 
\begin{eqnarray}
&& \hspace{-1cm} 
Q_{\Phi_{f^0}} (f^0) = 0 , 
\label{eq:null_space_Q}
\end{eqnarray}
i.e., borrowing to the terminology of statistical mechanics, $f^0$ is a Local Thermodynamical Equilibrium (LTE).  Therefore, we need to determine the set of LTE's of the collision operator. 

\begin{figure}[htbp]
\begin{center}
\input{Nash_2.pstex_t}
\caption{The LTE distribution $u \in {\mathbb S}^1 \to M_\Phi(u,\xi)$ for a given target point $\xi \in {\mathbb R}^2$ as a function of $u$ in polar coordinates (blue curve). The distribution $M_\Phi$ and the potential $\Phi$ (black dashed curve) have opposite monotonies.  The maxima of $M_\Phi$ are indicated by black semi-lines. Their width are roughly proportional to $\sqrt d$. The direction of the mean velocity $U$ is indicated by the red semi-line. It is fully determined by $M_\Phi$ and therefore, by $\Phi$ and is a function of $(x,\xi,t)$. We have $|U| < 1$. }
\label{fig:Nash}
\end{center}
\end{figure}
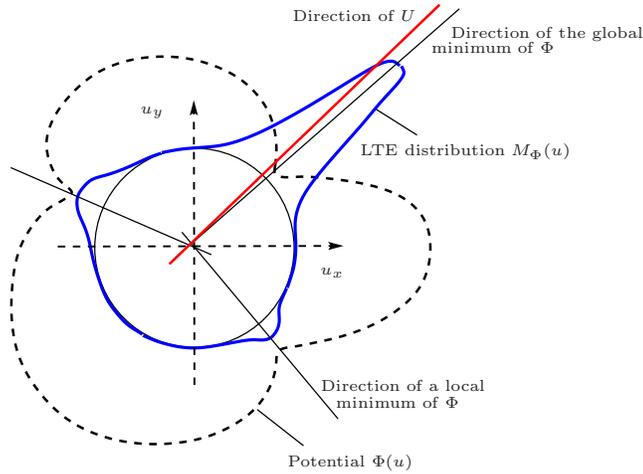

Let us first assume that the function $\Phi$: $(u,\xi) \in {\mathbb S}^1 \times {\mathbb R}^2 \to \Phi(u,\xi) \in {\mathbb R}$ is a given function. We introduce: 
\begin{eqnarray}
&& \hspace{-1cm} 
M_\Phi(u,\xi) = \frac{1}{Z_\Phi(\xi)} \exp \big( - \frac{ \Phi_\Phi(u,\xi) }{d} \big) .   
\label{eq:hydro_LED}
\end{eqnarray}
The quantity $Z_\Phi(\xi)$ is the normalizing constant, i.e. is such that 
\begin{eqnarray}
&& \hspace{-1cm} 
\int_{u \in {\mathbb S}^1} M_\Phi(u,\xi) \, du = 1, \quad \quad \forall \xi \in {\mathbb R}^2. 
\label{eq:hydro_LED_normal}
\end{eqnarray}
We provide a graphical representation of the function $u \in {\mathbb S}^1 \to M_\Phi(u,\xi)$ for a given $\xi \in {\mathbb S}^1$ in Fig.~\ref{fig:Nash} (blue curve) in polar coordinates. We realize that $M_\Phi$ and the potential $\Phi$ (black dashed curve) have opposite monotonies, as they should, given (\ref{eq:hydro_LED}). The noise intensity $d$ characterizes the width of the maxima of $M_\Phi$.

With (\ref{eq:hydro_Q}), the collision operator can be written:
\begin{eqnarray}
&& \hspace{-1cm} 
Q_\Phi(f) = -  d \, \nabla_u \cdot \Big( M_\Phi \nabla_u \big( \frac{f}{M_\Phi} \big) \Big). 
\label{eq:hydro_Q_2}
\end{eqnarray}
As a consequence of Green's formula, for any function $f(u,\xi)$ with appropriate regularity, we have:
\begin{eqnarray}
&& \hspace{-1cm} 
\int_{(u,\xi) \in {\mathbb S}^1 \times {\mathbb R}^2} Q_\Phi(f) \, \frac{f}{M_\phi} \, du \, d\xi =  
- \int_{(u,\xi) \in {\mathbb S}^1 \times {\mathbb R}^2} M_\Phi \, \Big|  \nabla_u \big( \frac{f}{M_\Phi} \big) \Big|^2 \, du \, d\xi.  
\label{eq:hydro_intQ}
\end{eqnarray}
Then, $f$ is a solution of the equation 
\begin{eqnarray}
&& \hspace{-1cm} 
Q_\Phi(f) = 0 , 
\label{eq:hydro_nullQ_1}
\end{eqnarray}
if and only if there exists a function $\rho$: $\xi  \in {\mathbb R}^2 \to \rho(\xi) \geq 0$ such that 
\begin{eqnarray}
f(u,\xi) = \rho(\xi) \, M_\Phi(u,\xi). 
\label{eq:hydro_nullQ_2}
\end{eqnarray}
The proof is analogous to that of formula (4.51) of \cite{Degond_etal_Heuristics} and is omitted. 

Therefore, an LTE is necessarily of the form (\ref{eq:hydro_nullQ_2}). However, for a given function $\rho$: $\xi  \in {\mathbb R}^2 \to \rho(\xi) \geq 0$, not all potentials $\Phi$ are allowed. Indeed, $f = \rho M_\Phi$ is a solution of (\ref{eq:null_space_Q}) if and only if we can ensure that $\Phi=\Phi_f$. In view of (\ref{eq:hydro_nullQ_2}) and (\ref{eq:hydro_Phi_local}), this constraint is written: 
\begin{eqnarray}
&& \hspace{-1cm} \Phi(u,\xi) = \Phi_t(u,\xi) - \frac{ \int_{(v,\eta) \in {\mathbb S}^1 \times {\mathbb R}^2}   \Psi(|v-u|) \, \rho(\eta) \, M_\Phi(v,\eta) \, dv \, d\eta }
{\int_{\eta \in {\mathbb R}^2}  \, \rho(\eta) \, d\eta }. 
\label{eq:hydro_Drho}
\end{eqnarray}
This is a functional equation for $\Phi$. It allows to find $\Phi$ as a functional of $\rho$. For the time being, we leave aside the question of the existence and uniqueness of solutions of this functional equation and admit that there is at least one isolated branch of solutions denoted by $\Phi_{[\rho]}$. Then, the LTE's are of the form $\rho \, M_{\Phi_{[\rho]}}$. 

Now, we restore the dependence upon $(x,t)$. The functions $\xi \to \rho(\xi)$ are parametrized by $(x,t)$ and denoted by $\rho_{(x,t)}$. The functional equation (\ref{eq:hydro_Drho}) is also parametrized by $x$, through the dependence of the function $\Phi_t$ upon $x$ (see (\ref{eq:Phi_t})) and its solutions are denoted by $\Phi_{x,[\rho]}$. Then,  the solutions of (\ref{eq:null_space_Q}) are given by:
\begin{eqnarray}
f^0(x,u,\xi,t) = \rho_{(x,t)}(\xi) \, M_{\Phi_{x,[\rho_{(x,t)}]}}(u,\xi), 
\label{eq:hydro_LED_2}
\end{eqnarray}
where, for any $(x,t)$, the function $(u,\xi) \to \Phi_{x,[\rho_{(x,t)}]}(u,\xi)$ satisfies (\ref{eq:hydro_Drho}). Thanks to the normalization condition (\ref{eq:hydro_LED_normal}), $\rho_{(x,t)}(\xi)$ is the density of pedestrians at point $x$ and time $t$ with target point $\xi$. 

Now, it remains to find the equations satisfied by the functions $\rho_{(x,t)}(\xi)$. First, we remark that 
\begin{eqnarray}
&& \hspace{-1cm} 
\int_{u \in {\mathbb S}^1} Q_{\Phi_f}(f) \, du = 0. 
\label{eq:hydro_intQ=0}
\end{eqnarray}
Consequently, if we integrate (\ref{eq:hydro_f}) with respect to $u$ and use (\ref{eq:hydro_intQ=0}), we find:
\begin{eqnarray}
&& \hspace{-1cm} 
\partial_t \rho^\varepsilon + \nabla_x \cdot (c \rho^\varepsilon U^\varepsilon) = 0. 
\label{eq:hydro_rho}
\end{eqnarray}
The functions $\rho^\varepsilon(x,\xi,t)$ and $c U^\varepsilon(x,\xi,t)$ are respectively the density and mean velocity of pedestrians at position $x$, time $t$ and target point $\xi$. They are given by:
\begin{eqnarray}
&& \hspace{-1cm} 
\rho^\varepsilon(x,\xi,t) =  \int_{u \in {\mathbb S}^1} f^\varepsilon(x,u,\xi,t) \, du , \quad (\rho^\varepsilon U^\varepsilon)(x,\xi,t) = \int_{u \in {\mathbb S}^1} f^\varepsilon(x,u,\xi,t) \, u\, du. 
\label{eq:hydro_mom}
\end{eqnarray}
Eq. (\ref{eq:hydro_rho}) is the mass conservation equation for the pedestrians having the same target point $\xi$. Now, taking the limit $\varepsilon \to 0$ in (\ref{eq:hydro_mom}) and using (\ref{eq:convergence}) and (\ref{eq:hydro_LED_2}), we get
\begin{eqnarray}
&& \hspace{-1cm} 
\rho^\varepsilon(x,\xi,t)  \to \rho_{(x,t)}(\xi), \quad U^\varepsilon(x,\xi,t) \to U_{x,[\rho_{(x,t)}]}(\xi), 
\label{eq:hydro_mom_lim}
\end{eqnarray}
with 
\begin{eqnarray}
&& \hspace{-1cm} 
U_{x,[\rho]}(\xi) = \int_{u \in {\mathbb S}^1} M_{\Phi_{x,[\rho]}}(u,\xi) \, u\, du. 
\label{eq:hydro_vel}
\end{eqnarray}
Then, the limit $\varepsilon \to 0$ in (\ref{eq:hydro_rho}) leads to 
\begin{eqnarray}
&& \hspace{-1cm} 
\partial_t \rho_{(x,t)}(\xi) + \nabla_x \cdot (c \rho_{(x,t)}(\xi) U_{x,[\rho_{(x,t)}]}(\xi)) = 0.
\label{eq:hydro_rho_lim}
\end{eqnarray}

The hydrodynamic model consists of the continuity eq. (\ref{eq:hydro_rho_lim}) for the density $\rho_{(x,t)}(\xi)$ of pedestrians with target point $\xi$, supplemented by eq. (\ref{eq:hydro_vel}) giving the mean velocity $U_{x,[\rho]}(\xi)$ in terms of $\rho$ at any point $(x,t)$ in space-time. This relation is a functional one as it is expressed through the solution $\Phi_{x,[\rho]}$ of the functional equation (\ref{eq:hydro_Drho}). This functional equation couples all the target points $\xi$ altogether. It expresses that each pedestrian has found the optimal reaction taking into account both his target point and the reactions of the other pedestrians. This reaction is optimal in the sense that no pedestrian is able to make a better choice by acting on his own control variable only, namely his velocity $u$. In this sense, the solution of the functional equation corresponds to a Nash equilibrium, in a similar fashion as the hydrodynamic limit of the heuristic-based model of \cite{Moussaid_etal_PNAS11} proposed in \cite{Degond_etal_Heuristics}. This hydrodynamic model is a first-order model in the sense of the traffic literature, since the velocity is entirely known from the density. We emphasize that this model is spatially and temporally local, as the Nash equilibrium is realized at any point $x$ and at all times $t$. This model as well as that of \cite{Degond_etal_Heuristics} fits in the framework given in  \cite{Degond_etal_arXiv:1212.6130} which aims to relate game theory and kinetic theory. This analogy will be detailed in future work. 

The comparison between the model of 
\cite{Degond_etal_Heuristics} and the present one is developed in the next section.

%%%%%%%%%%%%%%%%%%%%%%%%%%%%%%%%%%%%%%%%%%%%%%%%
%%%%%%%%%%%%%%%%%%%%%%%%%%%%%%%%%%%%%%%%%%%%%%%%
%%%%%%%%%%%%%%%%%%%%%%%%%%%%%%%%%%%%%%%%%%%%%%%%
%%%%%%%%%%%%%%%%%%%%%%%%%%%%%%%%%%%%%%%%%%%%%%%%
\setcounter{equation}{0}
\section{Discussion}
\label{sec:discussion}

In this section, we mostly discuss the analogies and differences with \cite{Degond_etal_Heuristics}. Both \cite{Degond_etal_Heuristics} and the present paper propose models with similar general features and the comparison of \cite{Degond_etal_Heuristics} with the literature mostly applies unchanged to the present work. We refer the reader to \cite{Degond_etal_Heuristics} for this discussion.

In the present paper like in \cite{Degond_etal_Heuristics}, the same outline has been adopted. The major difference is in the way the elementary interactions between the pedestrians are conceived and incorporated in the models. They result from the fundamental differences between the IBM's of respectively \cite{Ondrej_etal_Siggraph10} (for the present work) and of \cite{Moussaid_etal_PNAS11} (for \cite{Degond_etal_Heuristics}). As already discussed at the end of section \ref{subsec_Nped}, the model of \cite{Ondrej_etal_Siggraph10} views the interaction between two pedestrians as a reaction to the threat of a collision and proposes a mechanistic view of this reaction. By contrast, the model of \cite{Moussaid_etal_PNAS11} proposes the vision of active agents performing rational choices in view of the satisfaction of a target. These differences lead to different rules in the definition of the bulk forces acting either on the kinetic or fluid models. In \cite{Degond_etal_Heuristics}, it was possible to express the optimization performed by the agents in the choice of their route as a potential-driven dynamics, with a suitable velocity potential. In \cite{Ondrej_etal_Siggraph10}, the action of the agents is dominated by a mechanistic view of their reaction which makes its expression in terms of potential-driven dynamics impossible. 

However, one of the merits of the present work is to propose a mild variant of the model of  \cite{Ondrej_etal_Siggraph10} where the vision of rational agents performing optimal choices could be restored, in an analogy with \cite{Guy_etal_PRE12, Hoogendoorn_Bovy_OptControlApplMeth03}. This modification is motivated by the analysis of certain configurations of the original model in which the reactions of the subjects seem unrealistic (see section \ref{subsec_Nped_modif}). With this modification, the IBM of \cite{Ondrej_etal_Siggraph10} could be made closer to  \cite{Moussaid_etal_PNAS11}. There are still differences in the way the agents perceive the scene and make decisions. In  \cite{Moussaid_etal_PNAS11}, the main sensor of the motion of the other pedestrians is the DTI. In \cite{Ondrej_etal_Siggraph10}, both the TTI (which is proportional to the DTI) and the DBA are used. Therefore, the model of \cite{Ondrej_etal_Siggraph10} is more elaborate in the perception phase. In \cite{Moussaid_etal_PNAS11}, the decision-making is performed by minimizing the distance to the target, subject to the constraint of no-collision. In the potential-driven modification of the model of \cite{Ondrej_etal_Siggraph10}, the decision-making is based on a cost function which combines the satisfaction of the target and the collision avoidance constraint in a more balanced way. 

In spite of these differences, the two models bear strong analogies, which is reflected in the analogies that can be noticed at the level of the fluid equations. Indeed, in the case of the monokinetic and VMF closures, the general structures of the models issued from \cite{Moussaid_etal_PNAS11} and \cite{Ondrej_etal_Siggraph10} are the same and the differences appear only in the details of the computation of the fluid force. In the case of the hydrodynamic limit, this analogy is even stronger, since both models rely on the resolution of a fixed point equation which translates the search for a Nash equilibrium. Again, the details of the computations of these equilibria are different, since different sensors of the collision are used. In particular, it seems that the hydrodynamic limit presented here is slightly simpler than that of \cite{Degond_etal_Heuristics}. It also involves more details of the interaction dynamics and seems likely to provide better results. However, the difference is tenuous and the basic principles of the two models are similar. Numerical comparisons between the two models and the experimental data should be able to decide which of the two models is the most efficient.

%%%%%%%%%%%%%%%%%%%%%%%%%%%%%%%%%%%%%%%%%%%%%%%%
%%%%%%%%%%%%%%%%%%%%%%%%%%%%%%%%%%%%%%%%%%%%%%%%
%%%%%%%%%%%%%%%%%%%%%%%%%%%%%%%%%%%%%%%%%%%%%%%%
%%%%%%%%%%%%%%%%%%%%%%%%%%%%%%%%%%%%%%%%%%%%%%%%
\setcounter{equation}{0}
\section{Conclusion}
\label{sec:conclu}

In this article, we have derived a hierarchy of continuum crowd dynamic models from the Individual-Based Model of \cite{Ondrej_etal_Siggraph10}. This IBM relies on a vision-based framework: the pedestrians analyze the scene and react to the collision threatening partners by changing their direction of motion, while trying to keep their target. We have first proposed a kinetic version of this IBM. Then, three types of fluid models are derived from the kinetic formulation. They are respectively associated to a monokinetic closure, a von Mises-Fisher closure and a hydrodynamic limit. These models are, to the best of our knowledge, the first macroscopic pedestrian models based on a microscopic vision-based models. In future work, numerical simulations will be developed to assess the validity of the model and compare it to other models (such as in \cite{Degond_etal_Heuristics}) and to experimental data.

%%%%%%%%%%%%%%%%%%%%%%%%%%%%%%%%%%%%%%%%%%%%%%%%
%%%%%%%%%%%%%%%%%%%%%%%%%%%%%%%%%%%%%%%%%%%%%%%%
%%%%%%%%%%%%%%%%%%%%%%%%%%%%%%%%%%%%%%%%%%%%%%%%
%%%%%%%%%%%%%%%%%%%%%%%%%%%%%%%%%%%%%%%%%%%%%%%%

%%%%%%%%%%%%%%%%%%%%%%%%%%%%%%%%%%%%%%%%%%%%%%%%%%%%%%%%%%%%%%%%%%%%%%%%%%%%%%%%%%%%%%%%%%%%%%%%
\end{document}

%% file: collision_1_1_2.pstex_t
\begin{picture}(0,0)%
\includegraphics{collision_1_1_2.pstex}%
\end{picture}%
\setlength{\unitlength}{2072sp}%
\begingroup\makeatletter\ifx\SetFigFont\undefined%
\gdef\SetFigFont#1#2#3#4#5{%
  \reset@font\fontsize{#1}{#2pt}%
  \fontfamily{#3}\fontseries{#4}\fontshape{#5}%
  \selectfont}%
\fi\endgroup%
\begin{picture}(7209,5518)(-1148,-4338)
\put(2371,-1471){\makebox(0,0)[lb]{\smash{{\SetFigFont{8}{9.6}{\familydefault}{\mddefault}{\updefault}{\color[rgb]{0,0,0}$\alpha_{ij}$}%
}}}}
\put(1434,-2148){\makebox(0,0)[lb]{\smash{{\SetFigFont{8}{9.6}{\familydefault}{\mddefault}{\updefault}{\color[rgb]{0,0,0}$v_i$}%
}}}}
\put( 94,-1018){\makebox(0,0)[lb]{\smash{{\SetFigFont{8}{9.6}{\familydefault}{\mddefault}{\updefault}{\color[rgb]{0,0,0}$u_i^\bot$}%
}}}}
\put(-114,-1906){\makebox(0,0)[lb]{\smash{{\SetFigFont{8}{9.6}{\familydefault}{\mddefault}{\updefault}{\color[rgb]{0,0,0}$x_i$}%
}}}}
\put(754,-2128){\makebox(0,0)[lb]{\smash{{\SetFigFont{8}{9.6}{\familydefault}{\mddefault}{\updefault}{\color[rgb]{0,0,0}$u_i$}%
}}}}
\put(2551,-1181){\makebox(0,0)[lb]{\smash{{\SetFigFont{8}{9.6}{\familydefault}{\mddefault}{\updefault}{\color[rgb]{0,0,0}$v_j - v_i$}%
}}}}
\put(3808,-914){\makebox(0,0)[lb]{\smash{{\SetFigFont{8}{9.6}{\familydefault}{\mddefault}{\updefault}{\color[rgb]{0,0,0}$v_j$}%
}}}}
\put(3679,321){\makebox(0,0)[lb]{\smash{{\SetFigFont{8}{9.6}{\familydefault}{\mddefault}{\updefault}{\color[rgb]{0,0,0}$x_j$}%
}}}}
\put(4891,279){\makebox(0,0)[lb]{\smash{{\SetFigFont{8}{9.6}{\familydefault}{\mddefault}{\updefault}{\color[rgb]{0,0,0}$k_{ij}$}%
}}}}
\put(4071,949){\makebox(0,0)[lb]{\smash{{\SetFigFont{8}{9.6}{\familydefault}{\mddefault}{\updefault}{\color[rgb]{0,0,0}$k_{ij}^\bot$}%
}}}}
\put(3917,-3075){\makebox(0,0)[lb]{\smash{{\SetFigFont{8}{9.6}{\rmdefault}{\mddefault}{\updefault}{\color[rgb]{0,0,0}divided by $|v_j-v_i|$}%
}}}}
\put(3877,-2795){\makebox(0,0)[lb]{\smash{{\SetFigFont{8}{9.6}{\rmdefault}{\mddefault}{\updefault}{\color[rgb]{0,0,0}TTI = this distance}%
}}}}
\put(-763,-2635){\makebox(0,0)[lb]{\smash{{\SetFigFont{8}{9.6}{\rmdefault}{\mddefault}{\updefault}{\color[rgb]{0,0,0}MD =}%
}}}}
\put(-1133,-2915){\makebox(0,0)[lb]{\smash{{\SetFigFont{8}{9.6}{\rmdefault}{\mddefault}{\updefault}{\color[rgb]{0,0,0}this distance}%
}}}}
\end{picture}%

%% file: Phi.pstex_t
\begin{picture}(0,0)%
\includegraphics{Phi.pstex}%
\end{picture}%
\setlength{\unitlength}{2072sp}%
\begingroup\makeatletter\ifx\SetFigFont\undefined%
\gdef\SetFigFont#1#2#3#4#5{%
  \reset@font\fontsize{#1}{#2pt}%
  \fontfamily{#3}\fontseries{#4}\fontshape{#5}%
  \selectfont}%
\fi\endgroup%
\begin{picture}(6791,5293)(13,-4826)
\put(6511,-3071){\makebox(0,0)[lb]{\smash{{\SetFigFont{9}{10.8}{\rmdefault}{\mddefault}{\updefault}{\color[rgb]{0,0,0}$\tau$}%
}}}}
\put(2471,109){\makebox(0,0)[lb]{\smash{{\SetFigFont{9}{10.8}{\rmdefault}{\mddefault}{\updefault}{\color[rgb]{0,0,0}$\Phi$}%
}}}}
\put(3961,-1851){\makebox(0,0)[lb]{\smash{{\SetFigFont{9}{10.8}{\rmdefault}{\mddefault}{\updefault}{\color[rgb]{0,0,0}$\Phi = \Phi_0 \sigma(\tau)$}%
}}}}
\put(3041,-461){\makebox(0,0)[lb]{\smash{{\SetFigFont{9}{10.8}{\rmdefault}{\mddefault}{\updefault}{\color[rgb]{0,0,0}$\Phi = \Phi_0 \sigma_0$}%
}}}}
\put(1521,-4391){\makebox(0,0)[lb]{\smash{{\SetFigFont{9}{10.8}{\rmdefault}{\mddefault}{\updefault}{\color[rgb]{0,0,0}$|\dot \alpha| = \sigma_0$}%
}}}}
\put(4591,-3501){\makebox(0,0)[lb]{\smash{{\SetFigFont{9}{10.8}{\rmdefault}{\mddefault}{\updefault}{\color[rgb]{0,0,0}$|\dot \alpha| = \sigma(\tau)$}%
}}}}
\put(431,-4721){\makebox(0,0)[lb]{\smash{{\SetFigFont{9}{10.8}{\rmdefault}{\mddefault}{\updefault}{\color[rgb]{0,0,0}$|\dot \alpha|$}%
}}}}
\put(2581,-3381){\makebox(0,0)[lb]{\smash{{\SetFigFont{9}{10.8}{\rmdefault}{\mddefault}{\updefault}{\color[rgb]{0,0,0}$\Phi = 0$}%
}}}}
\end{picture}%

%% file: Psi.pstex_t
\begin{picture}(0,0)%
\includegraphics{Psi.pstex}%
\end{picture}%
\setlength{\unitlength}{2072sp}%
\begingroup\makeatletter\ifx\SetFigFont\undefined%
\gdef\SetFigFont#1#2#3#4#5{%
  \reset@font\fontsize{#1}{#2pt}%
  \fontfamily{#3}\fontseries{#4}\fontshape{#5}%
  \selectfont}%
\fi\endgroup%
\begin{picture}(7358,6199)(191,-5597)
\put(206,-5461){\makebox(0,0)[lb]{\smash{{\SetFigFont{12}{14.4}{\rmdefault}{\mddefault}{\updefault}$\hat {\mathcal S}_-(v-u)$}}}}
\put(926,-121){\makebox(0,0)[lb]{\smash{{\SetFigFont{12}{14.4}{\rmdefault}{\mddefault}{\updefault}$\hat {\mathcal S}_+(v-u)$}}}}
\put(5476,219){\makebox(0,0)[lb]{\smash{{\SetFigFont{12}{14.4}{\rmdefault}{\mddefault}{\updefault}$\zeta_\bot$}}}}
\put(7091,-2946){\makebox(0,0)[lb]{\smash{{\SetFigFont{12}{14.4}{\rmdefault}{\mddefault}{\updefault}$\zeta_\parallel$}}}}
\put(5476,-2896){\makebox(0,0)[lb]{\smash{{\SetFigFont{12}{14.4}{\rmdefault}{\mddefault}{\updefault}$v-u$}}}}
\put(6326,-3861){\makebox(0,0)[lb]{\smash{{\SetFigFont{12}{14.4}{\rmdefault}{\mddefault}{\updefault}$\widehat{ \dot \alpha} =$ Constant $<0$}}}}
\put(6276,-1611){\makebox(0,0)[lb]{\smash{{\SetFigFont{12}{14.4}{\rmdefault}{\mddefault}{\updefault}$\widehat{ \dot \alpha} =$ Constant $>0$}}}}
\put(2286,-5176){\makebox(0,0)[lb]{\smash{{\SetFigFont{12}{14.4}{\rmdefault}{\mddefault}{\updefault}$|\hat \tau| =$ Constant}}}}
\end{picture}%

%% file: VMF.pstex_t
\begin{picture}(0,0)%
\includegraphics{VMF.pstex}%
\end{picture}%
\setlength{\unitlength}{1243sp}%
\begingroup\makeatletter\ifx\SetFigFont\undefined%
\gdef\SetFigFont#1#2#3#4#5{%
  \reset@font\fontsize{#1}{#2pt}%
  \fontfamily{#3}\fontseries{#4}\fontshape{#5}%
  \selectfont}%
\fi\endgroup%
\begin{picture}(10577,8192)(1952,-5820)
\put(9541,2069){\makebox(0,0)[lb]{\smash{{\SetFigFont{7}{8.4}{\rmdefault}{\mddefault}{\updefault}Direction of $U$}}}}
\put(9286,-1621){\makebox(0,0)[lb]{\smash{{\SetFigFont{7}{8.4}{\rmdefault}{\mddefault}{\updefault}VMF distribution $M_{U}(u)$}}}}
\put(5731,1514){\makebox(0,0)[lb]{\smash{{\SetFigFont{7}{8.4}{\rmdefault}{\mddefault}{\updefault}Dirac delta distribution $\delta_{U}(u)$}}}}
\put(3976,719){\makebox(0,0)[lb]{\smash{{\SetFigFont{7}{8.4}{\rmdefault}{\mddefault}{\updefault}$u_y$}}}}
\put(8281,-3151){\makebox(0,0)[lb]{\smash{{\SetFigFont{7}{8.4}{\rmdefault}{\mddefault}{\updefault}$u_x$}}}}
\end{picture}%

%% file: Nash_2.pstex_t
\begin{picture}(0,0)%
\includegraphics{Nash_2.pstex}%
\end{picture}%
\setlength{\unitlength}{1036sp}%
\begingroup\makeatletter\ifx\SetFigFont\undefined%
\gdef\SetFigFont#1#2#3#4#5{%
  \reset@font\fontsize{#1}{#2pt}%
  \fontfamily{#3}\fontseries{#4}\fontshape{#5}%
  \selectfont}%
\fi\endgroup%
\begin{picture}(14184,11245)(831,-7862)
\put(7621,2939){\makebox(0,0)[lb]{\smash{{\SetFigFont{6}{7.2}{\rmdefault}{\mddefault}{\updefault}Direction of $U$}}}}
\put(3976,719){\makebox(0,0)[lb]{\smash{{\SetFigFont{6}{7.2}{\rmdefault}{\mddefault}{\updefault}$u_y$}}}}
\put(8281,-3151){\makebox(0,0)[lb]{\smash{{\SetFigFont{6}{7.2}{\rmdefault}{\mddefault}{\updefault}$u_x$}}}}
\put(9241,-286){\makebox(0,0)[lb]{\smash{{\SetFigFont{6}{7.2}{\rmdefault}{\mddefault}{\updefault}LTE distribution $M_\Phi(u)$}}}}
\put(7501,-7726){\makebox(0,0)[lb]{\smash{{\SetFigFont{6}{7.2}{\rmdefault}{\mddefault}{\updefault}Potential $\Phi(u)$}}}}
\put(11401,2519){\makebox(0,0)[lb]{\smash{{\SetFigFont{6}{7.2}{\rmdefault}{\mddefault}{\updefault}Direction of the global}}}}
\put(10936,2159){\makebox(0,0)[lb]{\smash{{\SetFigFont{6}{7.2}{\rmdefault}{\mddefault}{\updefault}minimum of $\Phi$}}}}
\put(8296,-5911){\makebox(0,0)[lb]{\smash{{\SetFigFont{6}{7.2}{\rmdefault}{\mddefault}{\updefault}Direction of a local}}}}
\put(8656,-6331){\makebox(0,0)[lb]{\smash{{\SetFigFont{6}{7.2}{\rmdefault}{\mddefault}{\updefault}minimum of $\Phi$}}}}
\end{picture}%